\documentclass[letterpaper,twocolumn,10pt]{article}

\PassOptionsToPackage{table}{xcolor}
\usepackage{usenix}

\usepackage{graphicx}
\usepackage{amssymb}
\usepackage{pifont}
\usepackage{makecell}

\usepackage{tikz}
\usetikzlibrary{mindmap, shadows}

\definecolor{ctiGray}{RGB}{82,82,82}        % center node

\definecolor{genMain}{RGB}{0,114,178}       % main branch
\definecolor{genSub}{RGB}{86,180,233}       % leaves

\definecolor{artMain}{RGB}{86,180,233}
\definecolor{artSub}{RGB}{136,204,238}

\definecolor{codiMain}{RGB}{230,159,0}
\definecolor{codiSub}{RGB}{253,190,80}

\definecolor{shareMain}{RGB}{213,94,0}
\definecolor{shareSub}{RGB}{244,143,66}

\definecolor{consMain}{RGB}{0,158,115}
\definecolor{consSub}{RGB}{120,198,121}
\definecolor{consLeaf}{RGB}{204,121,167}

\definecolor{centergray}{RGB}{82,82,82}
\definecolor{greenA}{RGB}{0,158,115}
\definecolor{greenB}{RGB}{0,180,125}
\definecolor{blueA}{RGB}{86,180,233}
\definecolor{blueB}{RGB}{100,200,255}
\definecolor{orangeA}{RGB}{230,159,0}
\definecolor{orangeB}{RGB}{255,190,60}
\definecolor{vermA}{RGB}{213,94,0}
\definecolor{vermB}{RGB}{230,120,60}
\definecolor{purpleA}{RGB}{204,121,167}
\definecolor{purpleB}{RGB}{225,145,190}
\definecolor{cbBlue}{RGB}{100,143,255}
\definecolor{cbLightBlue}{RGB}{120,170,255}
\definecolor{cbOrange}{RGB}{254,97,0}
\definecolor{cbLightOrange}{RGB}{255,131,43}
\definecolor{cbGreen}{RGB}{0,158,115}

\usepackage{amsmath}
\usepackage{enumitem}
\usepackage{adjustbox}
\usepackage{filecontents}
\usepackage{multirow}
\usepackage{float}

\usepackage{array}
\usepackage{booktabs}   % For better table rules
\usepackage{rotating}   % For rotating text
\usepackage{longtable}
\usepackage{tcolorbox}
\tcbuselibrary{breakable}

\definecolor{lightgray}{gray}{0.9}
\definecolor{lightergray}{gray}{0.95}
\usepackage{xspace}

\newcommand{\mypar}[1]{\vspace{3px}

\noindent\textbf{#1}\xspace}
\newcommand{\artifacturl}{\url{https://anonymous.4open.science/r/[placeholder]}\xspace}

\newcommand{\numberofquestions}{11\xspace}
\newcommand{\totalParticipants}{18\xspace}
\newcommand{\totalOrganizations}{5\xspace}
\newcommand{\totalParticipantsWord}{eighteen\xspace}
\newcommand{\totalOrganizationsWord}{five\xspace}

\newcommand{\filteredPapersNum}{123\xspace}
\newcommand{\filteredCtiConsumptionPapers}{71\xspace}
\newcommand{\filteredThreatDataPapers}{44\xspace}
\newcommand{\filteredCtiGenSharingPapers}{16\xspace}

\newcommand{\surveyExtractionChallengeNum}{13\xspace}
\newcommand{\surveyTtpChallengeNum}{10\xspace}
\newcommand{\surveyRedactionChallengeNum}{17\xspace}
\newcommand{\surveyFormatChallengeNum}{7\xspace}
\newcommand{\surveyAiCostHurdleNum}{6\xspace}
\newcommand{\surveyAiPrivacyHurdleNum}{13\xspace}
\newcommand{\surveySharingPlatformNum}{10\xspace}

\newcommand{\modelPropA}{Claude Opus 4.8\xspace}
\newcommand{\modelPropB}{GPT-5.5\xspace}
\newcommand{\modelOpen}{GLM 5.2\xspace}

\newcommand{\evalModelsNum}{3\xspace}
\newcommand{\evalCampaignsNum}{5\xspace}
\newcommand{\evalSamplesNum}{49\xspace}
\newcommand{\evalTuningReportsNum}{4\xspace}
\newcommand{\evalHeldoutReportsNum}{45\xspace}
\newcommand{\evalHeldoutCampaignsNum}{4\xspace}
\newcommand{\evalPlantedIdentifiersNum}{131\xspace}
\newcommand{\evalScoredIdentifiersNum}{113\xspace}
\newcommand{\evalHeldoutIdentifiersNum}{92\xspace}
\newcommand{\evalDescriptivePhrasesNum}{10\xspace}

\newcommand{\evalCostTestPropA}{\$207\xspace}
\newcommand{\evalCostTestPropB}{\$84\xspace}
\newcommand{\evalCostTestOpen}{\$75\xspace}

\newcommand{\evalinsight}[1]{%
\begin{tcolorbox}[colback=lightergray,colframe=black!35,boxrule=0.4pt,arc=1pt,left=3pt,right=3pt,top=3pt,bottom=3pt]
\footnotesize\textbf{Insight.} #1
\end{tcolorbox}
}

\newtcolorbox{appendixbox}{%
breakable,
width=\columnwidth,
colback=lightergray,
colframe=black!35,
boxrule=0.4pt,
arc=1pt,
left=5pt,
right=5pt,
top=5pt,
bottom=5pt,
before skip=0.5\baselineskip,
after skip=0.75\baselineskip,
fontupper=\small\ttfamily,
before upper={\raggedright\setlength{\parindent}{0pt}\setlength{\parskip}{2pt}\setlength{\baselineskip}{1.15em}}%
}

\newtcolorbox{promptbox}{%
breakable,
width=\columnwidth,
colback=lightergray,
colframe=black!35,
boxrule=0.4pt,
arc=1pt,
left=4pt,
right=4pt,
top=4pt,
bottom=4pt,
before skip=0.5\baselineskip,
after skip=0.75\baselineskip,
fontupper=\scriptsize\ttfamily,
before upper={\raggedright\setlength{\parindent}{0pt}\setlength{\parskip}{2pt}\setlength{\baselineskip}{1.12em}}%
}

\begin{document}
\pagenumbering{arabic}
\date{}

\title{\Large \bf A SoK for SoCs: Reading the TI Leaves on AI for Cyber Threat Intelligence Generation and Sharing}

% \author{
% {\rm Saastha Vasan$^{1}$, Hadjer Benkraouda$^{2}$, Jizhou Chen$^{3}$, Leyan Pan$^{3}$, Doguhan Yeke$^{4}$, Shinan Liu$^{5}$,}\\
% {\rm Noah Spahn$^{1}$, Stefano Ortolani$^{6}$, David Evans$^{7}$, Christopher Kruegel$^{1,8}$, Giovanni Vigna$^{1,6}$}\\[4pt]
% {\rm $^{1}$UC Santa Barbara \quad $^{2}$University of Illinois at Urbana-Champaign \quad $^{3}$Georgia Institute of Technology}\\
% {\rm $^{4}$Purdue University \quad $^{5}$University of Chicago \quad $^{6}$Broadcom \quad $^{7}$University of Virginia \quad $^{8}$Cisco}\\[4pt]
% {\rm \small saastha@ucsb.edu, hadjerb2@illinois.edu, \{jzchen, leyanpan\}@gatech.edu, dyeke@purdue.edu, shinanliu@uchicago.edu,}\\
% {\rm \small ncs@ucsb.edu, stefano.ortolani@broadcom.com, evans@virginia.edu, chris@cs.ucsb.edu, vigna@ucsb.edu}
% }

%for single author (just remove % characters)
\author{
{\rm Saastha Vasan}\\
UC Santa Barbara\\
\and
{\rm Hadjer Benkraouda}\\
UIUC
\and
{\rm Jizhou Chen}\\
Georgia Tech
\and
{\rm Leyan Pan}\\
Georgia Tech
\and
{\rm Doguhan Yeke}\\
Purdue University
\and
{\rm Shinan Liu}\\
University of Chicago
\and
{\rm Noah Spahn}\\
UC Santa Barbara
\and
{\rm Stefano Ortolani}\\
Broadcom
\and
{\rm David Evans}\\
University of Virginia
\and
{\rm Christopher Kruegel}\\
UC Santa Barbara
\and
{\rm Giovanni Vigna}\\
UC Santa Barbara
}
% % copy the following lines to add more authors
% % \and
% % {\rm Name}\\
% %Name Institution
% } % end author

\maketitle

\begin{abstract}
Cyber Threat Intelligence (CTI) is essential for defending mission-critical infrastructure, yet the process of transforming raw attack evidence into shareable CTI remains fragmented and understudied.

We conduct a literature survey of \filteredPapersNum academic papers, organizing the CTI lifecycle into three stages: Threat Data Collection, CTI Generation and Sharing, and CTI Consumption.
The first and third stages are well represented in the literature, whereas only \filteredCtiGenSharingPapers{} papers address CTI Generation and Sharing.
To learn how this stage is practiced, we survey \totalParticipants{} practitioners across \totalOrganizations{} organizations who routinely generate and share CTI.
They describe a largely manual process with four recurring challenges: preventing the exposure of sensitive information, extracting indicators from noisy attack data, correlating observed behavior with standardized tactics, techniques, and procedures (TTPs), and translating CTI into the formats that sharing platforms require.

Using the insights from the practitioner survey, we divide the CTI Generation and Sharing stage into four steps: Intelligence Extraction, Normalization and Enrichment, Codification, and Distribution.
We then conduct pilot studies that probe the feasibility of current Large Language Models (LLMs) for each step.
The pilot studies show that LLMs can assist an analyst in each of the four steps.
However, the models recover only a fraction of the indicators the evidence contains, struggle to ground every claim in the supplied evidence, and do not judge what keeps shared intelligence useful to its recipients.
Each step therefore requires expert supervision.
Based on these observations, we derive three research directions for automating the production of shareable intelligence.
\end{abstract}

\section{Introduction} \label{section:introduction}

Cybersecurity professionals are charged with protecting their organizations against threat actors whose tactics, techniques, and procedures (TTPs) evolve continuously.
Keeping abreast of the latest malware artifacts, the methods used for their propagation, and the infrastructure used to control them is a daunting task~\cite{mahboubi2024evolving, vielberth2020security, botacin2021challenges}.
A central difficulty is that defenders have limited visibility into attacks that have not yet reached their own networks.
Although threat actors reuse attacks against multiple targets, each affected organization observes only a single instance of the threat and therefore lacks the broader context needed to mount an effective defense~\cite{liao2016acing, zhao2020cyber}.

Cyber Threat Intelligence (CTI) addresses this limitation through collaborative defense.
When one organization shares what it has learned from an attack, other organizations can detect and respond to the same threat~\cite{wagner2017relevance, dutta2020overview, NIST800150, shu2018threat, paladini2024you, zhu2018chainsmith, huang2017gossip, gao2021enabling, gao2021system, dekel2022mabat}.
However, such intelligence must first be produced from the evidence that an attack leaves behind.
Analysts must isolate the attacker's traces from noisy telemetry, describe the observed behavior in shared taxonomies, remove personally identifiable information (PII) and organizationally identifiable information (OII), and translate the result into the formats required by the sharing channels.
Errors at any of these steps are costly~\cite{griffioen2020quality}.
Incorrectly extracted indicators cause false positives for the recipients, and incomplete redaction exposes sensitive victim or organizational information.
A recent measurement of the threat intelligence ecosystem reflects the difficulty of this process, reporting that 67\% of security vendors analyze threats but only 17\% share intelligence~\cite{galloway2026actively}.

To understand how existing research covers this workflow, we conducted a literature survey of \filteredPapersNum{} papers spanning the CTI lifecycle.
The literature survey organizes the lifecycle into three stages, distinguished by the input: \emph{Threat Data Collection}, which transforms raw telemetry into local attack representations such as provenance graphs and sandbox reports~\cite{milajerdi2019holmes, alsaheel2021atlas, cuckoosandbox, hassan2020tactical, vasan2024deepcapa, ding2023airtag, jiang2025orthrus}; \emph{CTI Generation and Sharing}, which transforms these local representations into shareable intelligence; and \emph{CTI Consumption}, which validates, prioritizes, and applies intelligence shared by others~\cite{Buechel2025SoK, gao2023threatkg, liao2016acing}.
We found that while \emph{Threat Data Collection} and \emph{CTI Consumption} are well represented in the literature, with \filteredThreatDataPapers{} and \filteredCtiConsumptionPapers{} papers respectively, only \filteredCtiGenSharingPapers{} papers fall under \emph{CTI Generation and Sharing}.
Even within these \filteredCtiGenSharingPapers{} papers, the majority of them focus on evaluating threat intelligence that has already been shared or on establishing sharing standards, rather than on how the underlying threat intelligence is produced~\cite{sillaber2016data, li2019reading, bouwman2020different}.
The adjacent stages therefore define only the endpoints of this process, as \emph{Threat Data Collection} systems produce the attack data and \emph{CTI Consumption} systems expect intelligence already expressed in standards such as STIX~\cite{STIX21}.
How practitioners move from one to the other, however, is not recorded in the literature.

To learn how \emph{CTI Generation and Sharing} is practiced, we conducted a practitioner survey, recruiting \totalParticipantsWord{} participants across \totalOrganizationsWord{} organizations who routinely generate and share CTI.
The practitioner survey yielded three findings.
First, converting attack data into CTI remains largely manual, dominated by scripted data parsing and human analysis.
Only 5 of \totalParticipants{} participants use machine learning and 3 use Large Language Models (LLMs).
Second, four challenges recur: preventing the exposure of sensitive information during sharing (\surveyRedactionChallengeNum{} of \totalParticipants), isolating malicious indicators from noisy attack data (\surveyExtractionChallengeNum{} of \totalParticipants), mapping observed behavior to standard taxonomies (\surveyTtpChallengeNum{} of \totalParticipants), and translating CTI into different sharing formats (\surveyFormatChallengeNum{} of \totalParticipants).
Third, practitioners consider these same tasks the strongest candidates for support from artificial intelligence (AI), but they doubt its reliability and worry that it may compromise data privacy.
Drawing on the literature survey and the practitioner survey, we structure \emph{CTI Generation and Sharing} into four steps, each defined by one of these challenges: \emph{Intelligence Extraction}, \emph{Normalization and Enrichment}, \emph{Codification}, and \emph{Distribution}.

These four steps involve substantial language processing, in which analysts interpret heterogeneous logs and restructure the content relevant to security according to shared standards and formats.
LLMs suit such tasks, as they are agnostic to input and output formats and apply broad knowledge of operating systems, tools, and attacker behavior directly from task instructions.
However, prior work applies natural language processing (NLP) and LLMs to \emph{CTI Consumption}, extracting indicators and techniques from published threat reports~\cite{husari2017ttpdrill, rani2023ttphunter, naveen2020deep, mischinger2024ioc}.
The trained extractors that do start from raw evidence are bound to one input format and to their training data, and labeled data for the generation steps is scarce~\cite{vasan2024deepcapa, sajid2021soda}.
Whether LLMs can support producing intelligence from an organization's own evidence remains an open question.

We therefore conducted four pilot studies, one for each step, with three recent LLMs of comparable capability, the proprietary \modelPropA{} and \modelPropB{} and the open-weight \modelOpen{}.
The first two pilot studies test whether a model can distinguish malicious indicators of compromise (IoCs) from benign noise and map the observed malicious behavior to MITRE ATT\&CK~\cite{mitre_attack} techniques.
The third study tests whether a model can redact the PII and OII from a CTI report.
The fourth study tests whether a model can convert an analyst's findings into a STIX~2.1 file.

Existing datasets~\cite{alam2024ctibench, lange2024annoctr} do not provide the inputs and labels these pilot studies require.
Published threat reports are the finished product of an analyst who had access to the underlying evidence, and that evidence is almost never released.
Where raw data is available, it carries no labels marking which events matter.
We therefore built the evidence and the labels ourselves.
We executed \evalSamplesNum{} malware samples in a security vendor's instrumented sandbox and replayed \evalCampaignsNum{} real campaigns in a controlled network.
We also injected a victim organization's identity into the published incident report behind each campaign, an article from The DFIR Report~\cite{dfir}.
The initial labels came from the sandbox vendor's automatic annotations for the malware reports, and from the published reports together with our execution records for the campaigns.
Two analysts then developed the final labels for each experiment independently and resolved disagreements through discussion.

For the first two steps, Intelligence Extraction and Normalization and Enrichment, we found that the models separate malicious indicators from benign noise better than static rules and recover techniques that the rules fail to identify.
For PII and OII redaction, the models follow whichever goal their prompt states, privacy or utility, and find nearly every injected identifier when privacy comes first.
For Distribution, every model generates STIX files that preserve the supplied findings.
However, the models recover only a fraction of the labeled indicators, struggle to ground claims in the evidence, and do not judge what keeps a shared report useful.
Each step therefore requires an analyst's review.
We use these insights to derive three research directions that we develop in \autoref{sec:discussion}: agentic workflows that decompose the generation steps and validate their own output, benchmarks that preserve every step of the generation pipeline, and redaction guided by explicit policies.

\mypar{Contributions.} Our key contributions are:
\begin{itemize}[noitemsep,topsep=0pt,leftmargin=*]

\item We conduct a comprehensive literature review of \filteredPapersNum{} CTI papers (\autoref{sec:lifecycle}), which enables us to formalize the CTI lifecycle into three stages.
We identify the \emph{CTI Generation and Sharing} phase as a significant research gap.

\item Motivated and informed by the findings of our literature review, we conducted a discovery-driven practitioner survey and report results from our survey, including insights about factors limiting adoption of LLMs (\autoref{sec:practitioner_survey}).

\item Leveraging insights from both the survey and the literature, we develop a structured framework for \emph{CTI Generation and Sharing}, defining the procedural steps required to transform raw evidence into shareable intelligence (\autoref{sec:cti_generation_and_sharing}).

\item We perform an exploratory evaluation of LLMs across four experiments aligned with the framework steps (intelligence extraction, normalization and enrichment, codification, and distribution) to assess the potential and limitations of AI-driven automation in CTI (\autoref{sec:evaluation}).

\item We identify barriers to automation, including recurring difficulty grounding technique mappings in input evidence and distinguishing attacker from victim infrastructure during redaction, and outline future research directions to address them (\autoref{sec:discussion}).

\end{itemize}

\section{Literature Review and CTI Lifecycle}
\label{sec:lifecycle}\label{sec:litreview}

We first present a motivating example illustrating the transition from environment-specific attack artifacts to shareable threat intelligence.
We then describe the literature survey, use its findings to organize the CTI lifecycle into three stages, and analyze how existing research is distributed across them.

\begin{figure*}[t]
\centering
\includegraphics[trim=0cm 9.6cm 2.0cm 0cm, clip, width=0.7\textwidth]{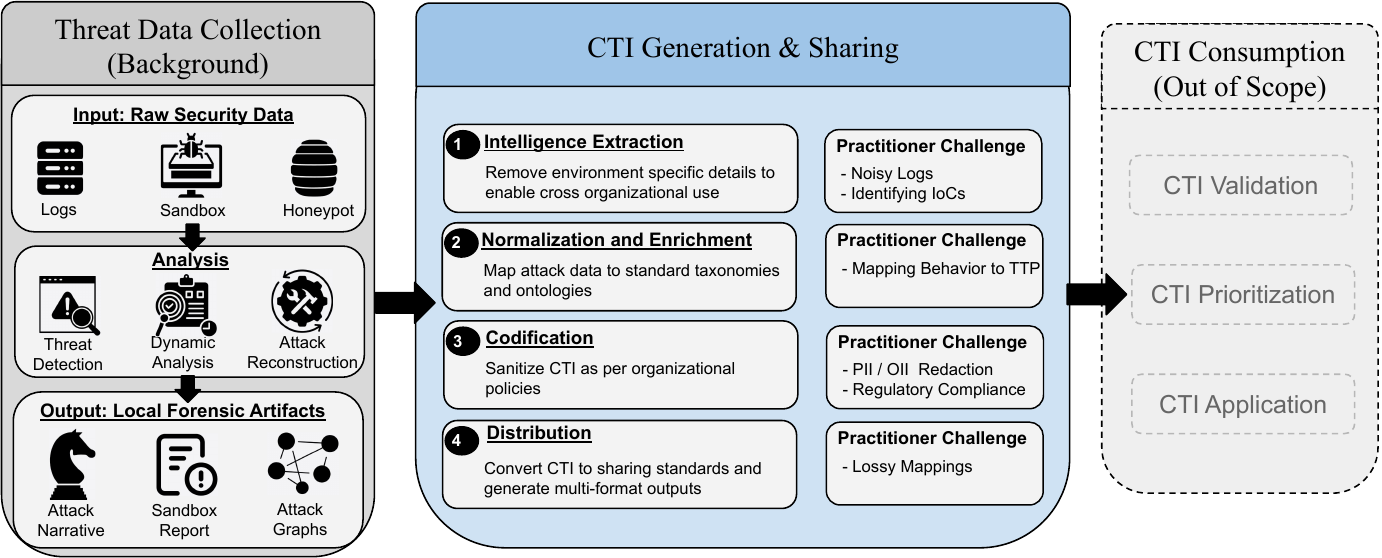}\caption{The CTI Lifecycle.
(1) \textbf{Threat Data Collection} takes raw telemetry as input and outputs attack data. (2) \textbf{CTI Generation and Sharing} transforms these local artifacts into shareable intelligence through four steps: Intelligence Extraction, Normalization and Enrichment, Codification, and Distribution.
The practitioner challenges at each step (shown on the right) are identified through the practitioner survey and discussed in Section~\ref{sec:practitioner_survey}.}
\label{fig:overview}
\end{figure*}

\mypar{Attack Scenario.}
An employee receives an email with a seemingly legitimate PDF attachment.
Upon opening the document, an embedded script triggers the download of a malware component from a remote server.
The malware establishes communication with a command-and-control (C\&C) server and begins scanning the internal network.
During reconnaissance, the malware identifies a database service vulnerable to CVE-2021-44228 (Log4Shell), exploits the flaw to gain unauthorized access, and exfiltrates gigabytes of sensitive data over an encrypted channel on an uncommon port.
This scenario illustrates how an attack progresses in stages from social engineering to exploitation and exfiltration, leaving behind heterogeneous traces across organizational defenses.

\mypar{Local Detection and Evidence.}
The organization's security stack detects the intrusion through three sensors: a sandbox identifies the malicious PDF, an intrusion detection system (IDS) flags the C\&C traffic, and an endpoint detection and response (EDR) component logs the internal scanning.
These systems perform \emph{Threat Data Collection} by isolating and structuring raw telemetry into attack data, such as provenance graphs or sandbox reports.
While this attack data provides a structured record of what happened, it is environment-specific and includes internal identifiers (e.g., local IP addresses and hostnames) and potentially sensitive information, making it unsuitable for immediate sharing.

\mypar{Intelligence Synthesis and Dissemination.}
To transform these local records into intelligence that other organizations can use (\emph{CTI Generation and Sharing}), analysts perform a series of refinement steps.
They extract artifacts, such as attacker-controlled IoCs and reusable behavioral patterns, from the local traces; map the observed behaviors to standard taxonomies and add contextual information such as actor attribution and known mitigations; redact sensitive organizational and personal information; and finally convert the result into the format required by the intended sharing channel.

\mypar{Collaborative Defense and Hunting.}
Another organization receives the shared CTI and validates its authenticity before assessing its relevance to its own infrastructure (\emph{CTI Consumption}).
To evaluate applicability, analysts check whether existing security controls can detect and mitigate the attack based on the provided TTPs.
Finally, they perform threat hunting to determine whether the threat is currently active.

This scenario illustrates the three stages of the CTI lifecycle that we define below.
The remainder of this paper focuses primarily on the middle stage, \emph{CTI Generation and Sharing}.

\subsection{Literature Corpus Construction}
The corpus of the literature survey covers the entire CTI lifecycle, from the systems that collect attack data to the works that describe systems and approaches that consume shared intelligence.

We searched two sources, DBLP and Google Scholar~\cite{googlescholar}, targeting publications from January 2000 through March 2026.
Keyword queries against DBLP retrieved 2,708 candidate entries.
After restricting these entries to established security and related venues and removing duplicates, we retained 153 candidate papers.
Queries against Google Scholar contributed 26 additional candidate papers after the same venue filtering and deduplication, for a joint candidate set of 179 papers.
The complete list of queries for both sources and the venue list are provided in Appendix~\ref{appendix:paper_selection}.
We then reviewed the title and abstract of each candidate for relevance to the CTI lifecycle, and this review resulted in 27 seed papers.
The reduction from 179 candidate papers to 27 reveals how CTI vocabulary is used in the literature.
For example, terms such as threat intelligence, TTP, and attack attribution appear in many abstracts as motivation for work whose contribution lies elsewhere, in areas such as adversarial machine learning, security operations, and Internet of Things (IoT) security.
Most keyword matches therefore mention CTI rather than study it.

At the same time, many papers that belong in the CTI lifecycle do not use CTI terminology in the title or abstract, the only text our candidate review examined.
For example, work on provenance-based forensics, dynamic malware analysis, and feed measurement rarely uses the phrase threat intelligence; keyword search alone therefore cannot retrieve these papers.
Starting from the seed papers, we therefore performed backward searches over their bibliographies and forward searches over citing works using Google Scholar, and these searches added 96 papers.
In total, the corpus comprises \filteredPapersNum{} primary papers.
We read each paper and classified it according to the lifecycle stage its primary contribution addresses, following the rule stated in the next subsection.
The complete classification, with the reasoning for every paper's placement, is provided in the released artifacts.

\subsection{CTI Lifecycle Framework}
\label{ssec:lifecycle_framework}
We organize the CTI lifecycle into three stages, illustrated in \autoref{fig:overview}: \emph{Threat Data Collection}, \emph{CTI Generation and Sharing}, and \emph{CTI Consumption}.

\mypar{Threat Data Collection.}
At this stage, raw telemetry, such as audit logs, application logs, and API call traces, is analyzed to reconstruct attack behavior.
The stage outputs structured representations of observed attacks, including attack timelines, attack graphs, behavior indicators, and malware sandbox reports.
Research in this domain concentrates on systems that perform provenance graph analysis, causal analysis, and malware sandbox report generation, described in \autoref{sec:threat_data_collection}.
However, the outputs of these systems remain bound to their analysis environments and include environment-specific identifiers that limit their direct utility outside the organization.

\mypar{CTI Generation and Sharing.}
This stage takes the localized outputs of \emph{Threat Data Collection} and generalizes them into CTI that can be shared across organizations.
The motivating example, together with the structure implied by the STIX~\cite{STIX21} sharing standard and the MITRE ATT\&CK~\cite{mitre_attack} taxonomy, outlines the work this stage performs.
Analysts perform the refinement steps described in the motivating example, from extraction through format translation, which Section~\ref{sec:cti_generation_and_sharing} examines in detail.

\mypar{CTI Consumption.}
This stage involves receiving shared CTI, validating it for correctness, prioritizing it based on organizational needs, and integrating it into defensive operations.
Research on this stage includes extraction from threat reports using NLP~\cite{Buechel2025SoK, husari2017ttpdrill}, knowledge graph construction~\cite{gao2023threatkg}, and feed quality evaluation~\cite{li2019reading, bouwman2020different}.
Recent systematizations provide detailed treatment of this stage~\cite{Buechel2025SoK, rahman2023what}.

\mypar{Classification Rule.}
We assign each paper to a stage by its input, not by its output.
A work that starts from intelligence that others already shared, such as published threat reports or feeds, belongs to \emph{CTI Consumption} even when its output is again a structured artifact such as an ATT\&CK mapping or a STIX bundle.
A work that starts from an organization's own attack data and produces intelligence belongs to \emph{CTI Generation and Sharing}.
This distinction matters in practice because many \emph{CTI Consumption} works emit the same standardized formats that \emph{CTI Generation and Sharing} works do, and the input is what separates the two.
A paper whose contributions genuinely span two stages receives both labels.

\subsection{CTI Lifecycle Coverage}

\begin{figure*}[t]
\centering
\resizebox{0.8\textwidth}{!}{%
% Taxonomy mindmap. Each leaf carries its paper count and the citations of its
% members; the per-paper classification rationale is in the paper
% classification appendix table. One \cite per bubble so the numbers sort and
% compress into ranges, and citation links are colored black inside the figure
% so they stay readable on the colored fills. Colors use three lightness bands
% so the branches remain distinguishable in grayscale.
\definecolor{rootgray}{RGB}{45,45,45}
\definecolor{tdcA}{RGB}{0,105,80}
\definecolor{tdcB}{RGB}{0,150,115}
\definecolor{gensA}{RGB}{205,125,0}
\definecolor{gensB}{RGB}{230,150,30}
\definecolor{consA}{RGB}{215,165,200}
\definecolor{consB}{RGB}{235,200,225}
\begingroup
\hypersetup{citecolor=black}
\begin{tikzpicture}[
  mindmap,
  every node/.style={
    concept,
    align=center,
    font=\bfseries\footnotesize
  },
  concept color=rootgray,
  level 1/.append style={
    level distance=4.4cm,
    minimum size=2.9cm,
    font=\bfseries\normalsize
  },
  level 2/.append style={
    level distance=4.0cm,
    minimum size=3.3cm,
    text width=3.0cm,
    font=\bfseries\small
  },
  tdc/.style={concept color=tdcA, every node/.append style={text=white}},
  tdcleaf/.style={concept color=tdcB, every node/.append style={text=black}},
  gens/.style={concept color=gensA, every node/.append style={text=black}},
  gensleaf/.style={concept color=gensB, every node/.append style={text=black}},
  cons/.style={concept color=consA, every node/.append style={text=black}},
  consleaf/.style={concept color=consB, every node/.append style={text=black}},
]

\node[minimum size=3cm, font=\bfseries\normalsize, text=white]{Cyber Threat\\Intelligence}
  % ===== 1. Threat Data Collection (lower right) =====
  child[tdc, grow=-90] {
    node {Threat Data\\Collection}
      child[tdcleaf, grow=-30]  { node {Organizational Logs\\{\footnotesize\mdseries(N=33)}\\{\scriptsize\mdseries\cite{alsaheel2021atlas,altinisik2023provg,bates2015trustworthy,cheng2024kairos,datta2022alastor,ding2023airtag,fang2022back,gehani2012spade,goyal2024r,han2020unicorn,hassan2019nodoze,hassan2020omegalog,hassan2020tactical,hossain2017sleuth,hossain2020combating,jia2024magic,jiang2025orthrus,king2003backtracking,king2005enriching,kwon2018mci,li2024nodlink,liu2018towards,m2018propatrol,ma2016protracer,milajerdi2019holmes,pei2016hercule,rehman2024flash,sun2025alerts,wang2022threatrace,xu2022depcomm,yang2020uiscope,yang2023prographer,zeng2021watson}}} }
      child[tdcleaf, grow=-150] { node {Sandbox Analysis\\{\footnotesize\mdseries(N=11)}\\{\scriptsize\mdseries\cite{Yin2007Panorama_Capturing,brengel2018memscrimper,dinaburg2008ether,inoue2008malware,kirat2014barecloud,lanzi2009ktracer,lengyel2014scalability,sajid2021soda,severi2018malrec,vasan2024deepcapa,willems2007toward}}} }
  }
  % ===== 2. CTI Generation and Sharing (lower left) =====
  child[gens, grow=180] {
    node {CTI\\Generation \&\\Sharing}
      child[gensleaf, grow=240] { node {Generation\\{\footnotesize\mdseries(N=7)}\\{\scriptsize\mdseries\cite{cabana2021threat,fereidooni2022fedcri,kurogome2019eiger,liao2016acing,sajid2021soda,sun2025alerts,vasan2024deepcapa}}} }
      child[gensleaf, grow=180] { node {Codification\\{\footnotesize\mdseries(N=3)}\\{\scriptsize\mdseries\cite{Shim2022OnDL,geras2024big,sillaber2016data}}} }
      child[gensleaf, grow=120] { node {Distribution\\{\footnotesize\mdseries(N=7)}\\{\scriptsize\mdseries\cite{Jin2024SharingCT,Wagner2016MISPTD,bouwman2022helping,geras2023sharing,purohit2020defensechain,sillaber2016data,stojkovski2021s}}} }
  }
  % ===== 3. CTI Consumption (top) =====
  child[cons, grow=0] {
    node {CTI\\Consumption}
      child[consleaf, grow=60]  { node {Validation\\{\footnotesize\mdseries(N=17)}\\{\scriptsize\mdseries\cite{Wagner2016MISPTD,bouwman2020different,bouwman2022helping,gao2018graph,geras2024big,jung_empirical_2004,kuhrer2014paint,li2019reading,liao2016acing,meier2018feedrank,pitsillidis_tasters_2012,qiang2018quality,sakellariou_methodology_2024,sheng_empirical_2009,sillaber2016data,wang2022comprehensive,zibak2022threat}}} }
      child[consleaf, grow=0]   { node {Application\\{\footnotesize\mdseries(N=52)}\\{\scriptsize\mdseries\cite{Buechel2025SoK,Rani2024TTPXHunterAT,abdeen2023smet,ahmed2024cyberentrel,alam2023looking,alves2022leveraging,ayoade2018automated,chen2025aecr,cheng2025ctinexus,dekel2022mabat,deliu2018collecting,dionisio2019cyberthreat,fayyazi2024advancing,fieblinger2024actionable,gao2021enabling,gao2021system,gao2023threatkg,ge2023explainable,ge2024seqmask,ghazi2018supervised,hao2024leveraging,huang2017gossip,huang2024mitretrieval,husari2017ttpdrill,husari2018using,kadoguchi2019exploring,kaiser2023attack,kim2022comparative,kumarasinghe2024semantic,legoy2020automated,liu2022threat,mischinger2024ioc,naveen2020deep,orbinato2022automatic,paladini2024you,park2023pretrained,rahman2024alert,rani2023ttphunter,satvat2021extractor,satyapanich2020casie,shu2018threat,siracusano2023time,tang2023attack,xu2024intelex,yan2022threat,you2022tim,you2024cyber,zhang2021ex,zhang2025attackg+,zhao2020cyber,zhao2020timiner,zhu2018chainsmith}}} }
      child[consleaf, grow=-60] { node {Prioritization\\{\footnotesize\mdseries(N=4)}\\{\scriptsize\mdseries\cite{de2023no,meier2018feedrank,sillaber2016data,wagner2017relevance}}} }
  };
\end{tikzpicture}
\endgroup
}
\caption{Systematization of \filteredPapersNum{} papers across the three stages of the CTI lifecycle: Threat Data Collection, CTI Generation and Sharing, and CTI Consumption.
The classification rationale for each paper is provided in the released artifacts (\artifacturl).}
\label{fig:cti_mindmap}
\end{figure*}
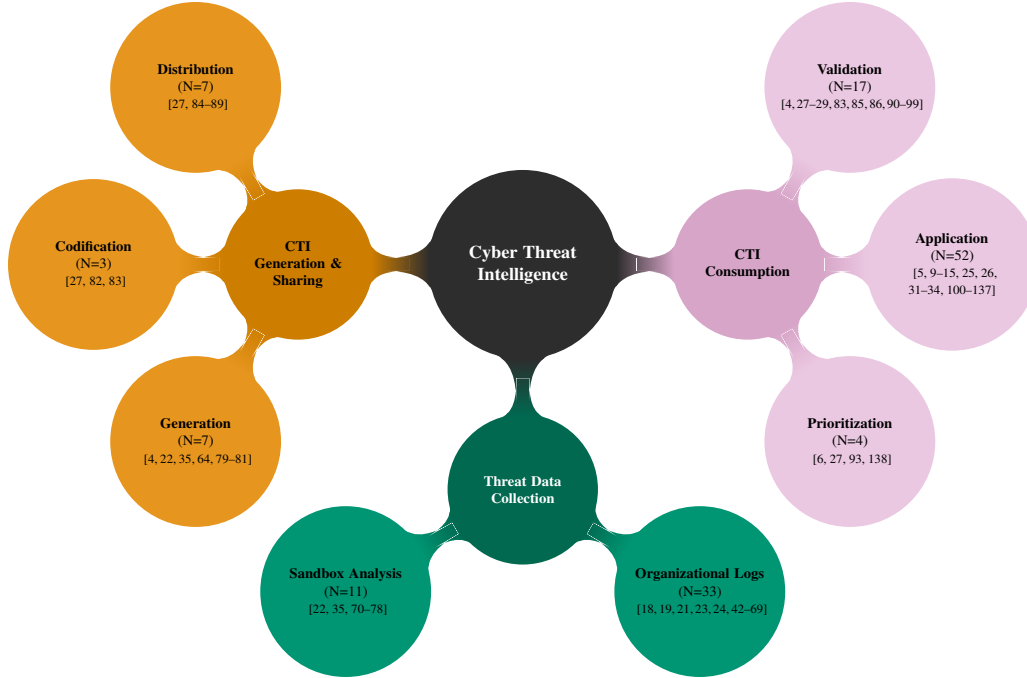

We now examine how existing research is distributed across the three stages.
\autoref{fig:cti_mindmap} shows how the \filteredPapersNum{} papers divide across the three stages under the classification rule stated above: \filteredThreatDataPapers{} papers fall under \emph{Threat Data Collection}, \filteredCtiGenSharingPapers{} under \emph{CTI Generation and Sharing}, and \filteredCtiConsumptionPapers{} under \emph{CTI Consumption}.
Eight papers belong to two stages.
SHIELD~\cite{sun2025alerts}, DeepCAPA~\cite{vasan2024deepcapa}, and SODA~\cite{sajid2021soda} span \emph{Threat Data Collection} and \emph{CTI Generation and Sharing}, as they produce local attack representations and also generate normalized, shareable intelligence.
Five additional works span \emph{CTI Generation and Sharing} and \emph{CTI Consumption}, as they combine sharing contributions with the validation or use of shared intelligence~\cite{Wagner2016MISPTD, sillaber2016data, geras2024big, liao2016acing, bouwman2022helping}.

The distribution reveals an imbalance.
\emph{Threat Data Collection} and \emph{CTI Consumption} are well represented in the literature, whereas only \filteredCtiGenSharingPapers{} of \filteredPapersNum{} papers address \emph{CTI Generation and Sharing}, and a recent active measurement of the threat intelligence ecosystem reports a corresponding imbalance in practice~\cite{galloway2026actively}.
Moreover, most of these \filteredCtiGenSharingPapers{} papers evaluate the quality of intelligence that has already been shared or establish sharing standards, rather than develop the methods to produce shareable intelligence in the first place.
The research community has invested in \textit{finding} threats and \textit{using} shared intelligence, while the process of \textit{producing} standardized, shareable intelligence from local evidence has received the least attention.

Recent systematizations and surveys address adjacent parts of the CTI lifecycle~\cite{inam2023sok, Buechel2025SoK, rahman2023what, furumoto2025comprehensive, mavroeidis2017cyber, alevizos2024towards}.
Several focus on provenance-based system auditing for \emph{Threat Data Collection}~\cite{inam2023sok} or on the extraction and processing of already shared CTI for \emph{CTI Consumption}~\cite{Buechel2025SoK, rahman2023what}, while measurement studies examine the quality, overlap, and propagation of shared CTI feeds~\cite{galloway2026actively, bouwman2020different, bouwman2022helping, furumoto2025comprehensive} and practitioner studies examine how analysts investigate and attribute threats~\cite{saha2025expert}.
Our work complements these efforts by describing the full lifecycle with emphasis on \emph{CTI Generation and Sharing}, grounding the analysis in the practitioner survey, and evaluating whether LLMs can support the steps of this stage.
This coverage also shapes the structure of the remainder of the paper.
Because \emph{CTI Consumption} is already systematized in detail~\cite{Buechel2025SoK, rahman2023what}, we do not devote a section to it.
\emph{Threat Data Collection} is likewise covered by existing systematizations~\cite{inam2023sok}, but its outputs are the inputs from which CTI is generated; \autoref{sec:threat_data_collection} therefore describes the forms this attack data takes before we examine how it is transformed.

The scarcity of literature on \emph{CTI Generation and Sharing} does not imply that the process is absent in practice; the practitioner survey (\autoref{sec:practitioner_survey}) confirms it is routinely performed.
Instead, the scarcity suggests that this work has historically been treated as a procedural, organizational task rather than a research problem.
This framing has left the process largely manual, hard to reproduce, and reliant on expert judgment.
Our evaluation in \autoref{sec:evaluation} shows that several steps of this stage present research challenges rather than engineering tasks.
Separating attacker activity from the benign system behavior that surrounds it in raw evidence requires a judgment that rules do not supply, redaction requires weighing what a detail reveals about the victim against its value to the recipient, and converting CTI into sharing formats can produce files that the receiving software rejects.

Investigating this stage, however, requires knowing how it is practiced.
The literature defines the stage's endpoints, the attack data that \emph{Threat Data Collection} systems produce and the shared intelligence that \emph{CTI Consumption} works expect.
It does not record which attack data practitioners begin with, which tools they rely on, which steps consume their effort, or whether they would delegate any of this work to automation.
To answer these questions, we conducted the practitioner survey presented in the next section.

\section{Practitioner Survey}
\label{sec:practitioner_survey}

To learn how \emph{CTI Generation and Sharing} is carried out in practice, we surveyed cybersecurity practitioners who routinely generate and share CTI.
The goal of the practitioner survey was threefold: (1) to determine the current state of automation in \emph{CTI Generation and Sharing}, (2) to identify the operational challenges practitioners face, and (3) to assess how practitioners view the potential of LLMs to address these challenges.
In this section, we describe the survey design, the participants, and the findings.

\subsection{Survey Design}
\label{ssec:survey_design}

\mypar{Recruitment and Demographics.}
We recruited \totalParticipants{} participants from \totalOrganizations{} organizations with dedicated threat intelligence teams.
Participants included SOC analysts, security researchers, and security engineers.
Eight of \totalParticipants{} participants reported more than five years of experience in cybersecurity; the sample therefore covers a range of experience levels.

\mypar{Questionnaire Structure.}
The questionnaire comprised \numberofquestions{} questions organized into three parts, one for each goal.
The first part asked which input data practitioners use to generate CTI and how they analyze it.
These questions establish which evidence the stage starts from and how automated its analysis currently is.
The second part asked which issues practitioners encounter while generating CTI from logs and while sharing the result, since the literature does not record where their difficulties concentrate.
The third part asked where practitioners see opportunities for AI in generating and sharing CTI and what hinders its adoption.
These questions capture the expectations that any proposal for automation must address.
The full set of questions and the aggregated responses are provided in \autoref{tab:survey_questions} in Appendix~\ref{appendix:survey_questionnaire}.
We did not collect or store participants' names or any other personal information.

\mypar{Intent and Bias Mitigation.}
To avoid steering the responses toward our own framing, we did not present participants with a predefined model of the \emph{CTI Generation and Sharing} process, and they described their workflows, difficulties, and expectations for automation independently.
The questionnaire used neutrally phrased multiple-choice questions in which practitioners could select one or more options, and every question included an Other category for open-ended answers; responses outside the listed options could therefore still be captured.

\subsection{Findings}
\label{ssec:survey_findings}

Although the participants came from \totalOrganizations{} different organizations, their responses concentrated on a small set of difficulties, which suggests that these difficulties are not specific to a single organization.
We organize the findings into three insights.

\mypar{Insight 1: Practitioners generate and share CTI with little automation.}
When asked for their main approaches to analyzing logs to generate CTI, practitioners most often selected scripted data parsing (15 of \totalParticipants) and manual analysis (13 of \totalParticipants), whereas only 5 of \totalParticipants{} selected machine learning and 3 of \totalParticipants{} LLMs.
These responses mirror the imbalance revealed by the literature survey.
Research has produced increasingly capable systems for collecting and structuring attack data, but the conversion of this data into shareable CTI has no comparable support.

\mypar{Insight 2: Four challenges recur across the process.}
Two questions asked which issues practitioners encounter, one about generating CTI from logs and one about sharing the resulting CTI.
The responses to both concentrate on four challenges.
The most widespread is ensuring that shared CTI does not expose private organizational or personal information (\surveyRedactionChallengeNum{} of \totalParticipants).
It is followed by noisy logs that prevent the extraction of relevant events (\surveyExtractionChallengeNum{} of \totalParticipants), the manual effort of correlating observed events with TTPs (\surveyTtpChallengeNum{} of \totalParticipants), and the translation of CTI into the formats that different sharing feeds require (\surveyFormatChallengeNum{} of \totalParticipants).
These four challenges define the four steps into which we structure the \emph{CTI Generation and Sharing} stage (\autoref{sec:cti_generation_and_sharing}).
The extraction of relevant events defines \emph{Intelligence Extraction}, the correlation with TTPs defines \emph{Normalization and Enrichment}, the protection of sensitive information defines \emph{Codification}, and format translation defines \emph{Distribution}.

\mypar{Insight 3: Practitioners consider the same tasks the best candidates for AI but doubt its reliability.}
When asked where AI could support generating and sharing CTI, practitioners most often selected the same tasks they had reported as challenges.
Filtering noisy logs to extract relevant events (15 of \totalParticipants) and mapping threat events to TTPs (13 of \totalParticipants) led the responses for generation, and finding and removing private information (14 of \totalParticipants) and translating CTI into different formats (11 of \totalParticipants) led the responses for sharing.
At the same time, practitioners raised two concerns that temper this interest.
The first concern is reliability.
When asked about extracting CTI from logs, 16 of \totalParticipants{} questioned whether AI could reliably filter raw logs, an option illustrated with errors in correlating and prioritizing events.
When asked about sharing, 14 of \totalParticipants{} reported a lack of confidence that AI systems would carry out the task correctly, and 13 of \totalParticipants{} expected them to introduce inaccuracies into the shared data.
The second concern is data privacy.
In the same question about sharing, 12 of \totalParticipants{} worried that insufficient controls could lead AI systems to break privacy or regulatory compliance.
When asked about the hurdles to incorporating AI into their workflow, \surveyAiPrivacyHurdleNum{} of \totalParticipants{} named data privacy.
These responses indicate an opportunity to reduce practitioner workload, provided that the reliability and privacy concerns are addressed.
Before evaluating whether current LLMs can perform these tasks, we describe the attack data that practitioners begin from (\autoref{sec:threat_data_collection}) and then examine each of the four steps in detail (\autoref{sec:cti_generation_and_sharing}).

\section{Threat Data Collection}
\label{sec:threat_data_collection}
Generating CTI begins with the evidence an organization already holds.
Before examining how that evidence becomes shareable intelligence, we describe the attack data available to analysts, since its form determines what the later steps can extract from it.
We group the data by its telemetry source.
Organizational logs record attacks as they unfold on production systems, and sandbox analysis records the behavior of malware executed in isolation.
Sandbox evidence is the most common starting point among the practitioner survey participants (17 of \totalParticipants).
In addition, 9 of \totalParticipants{} collect logs from active networks during incident forensics.
An industrial study and a recent independent re-evaluation report that the research systems described below fall short of operational requirements~\cite{dong2023we, bilot2025sometimes}.
This gap further motivates our focus on the evidence these systems produce rather than on the systems themselves.

\subsection{Attack Data from Organizational Logs}
\label{ssec:organizational_logs}

When an incident is detected, organizations reconstruct the adversary's trail from audit logs, application logs, and network telemetry.
Two decades of research on provenance tracking and causal analysis have produced systems that condense millions of low-level events into structured attack evidence, and their outputs take four forms.
Attack timelines reconstruct the sequence of steps behind an intrusion, tying each step to the processes, files, and network addresses involved~\cite{king2003backtracking, hossain2017sleuth, alsaheel2021atlas, ding2023airtag, pei2016hercule, ma2016protracer, datta2022alastor}.
Attack graphs represent these entities as nodes, capture their causal dependencies as edges, and typically attach a maliciousness score to each node~\cite{jiang2025orthrus, cheng2024kairos, li2024nodlink, rehman2024flash, jia2024magic, wang2022threatrace, goyal2024r, yang2023prographer, han2020unicorn, altinisik2023provg}.
Behavior indicators abstract low-level events into semantic descriptions of what an attack does.
WATSON~\cite{zeng2021watson} produces such descriptions from audit logs, and HOLMES~\cite{milajerdi2019holmes}, RapSheet~\cite{hassan2020tactical}, and SHIELD~\cite{sun2025alerts} express them in frameworks such as MITRE ATT\&CK.
Forensic paths reduce full dependency graphs to a small set of causal chains an analyst must review, countering the dependency explosion that makes raw provenance graphs impractical to inspect~\cite{liu2018towards, hassan2019nodoze, kwon2018mci, fang2022back, xu2022depcomm, king2005enriching, hossain2020combating, m2018propatrol, hassan2020omegalog, yang2020uiscope}.
Several of the systems above build on trustworthy provenance collection infrastructure~\cite{bates2015trustworthy, gehani2012spade}.
For \emph{CTI Generation and Sharing}, these outputs supply the entities, causal structure, and behavioral context from which indicators and techniques are extracted.

\subsection{Attack Data from Sandbox Analysis}
\label{ssec:sandbox_logs}

Sandbox analysis complements organizational telemetry with controlled and repeatable observation.
Executing a suspicious sample in isolation produces an execution log, a linear record of the sample's interaction with the operating system that covers process, file, registry, and network activity~\cite{willems2007toward, severi2018malrec, brengel2018memscrimper, inoue2008malware, lanzi2009ktracer, Yin2007Panorama_Capturing}.
Instrumentation frameworks capture this record at different levels of the software stack~\cite{lengyel2014scalability, dinaburg2008ether, song2008bitblaze}, hardened environments counter malware that detects analysis~\cite{kirat2014barecloud}, and open-source sandboxes such as Cuckoo~\cite{cuckoosandbox} and CAPE~\cite{cape} aggregate the raw events into more abstract observations.
Recent systems condense execution logs further into capability profiles, summaries of what a sample can do~\cite{sajid2021soda, vasan2024deepcapa}.
For \emph{CTI Generation and Sharing}, sandbox evidence reveals a sample's capabilities and the external infrastructure it contacts, without the background noise of a production network.

Despite the maturity of these systems, their outputs remain tied to the environments that produced them.
Timelines reference local hostnames, graphs embed internal addresses, and execution logs mix the sample's activity with artifacts of the analysis environment.
The next section examines how this locally bound evidence is transformed into shareable intelligence.

\section{CTI Generation and Sharing}
\label{sec:cti_generation_and_sharing}

\emph{CTI Generation and Sharing} takes as input the locally bound outputs of \emph{Threat Data Collection} described in \autoref{sec:threat_data_collection} (attack timelines, graphs, behavior indicators, forensic paths, and sandbox reports) and transforms them into intelligence that can move between organizations.
\emph{CTI Consumption} systems build on its output.
Research in that stage extracts indicators and techniques from published threat reports~\cite{husari2017ttpdrill, legoy2020automated, satvat2021extractor, gao2023threatkg} and measures the quality of shared feeds~\cite{li2019reading, bouwman2020different}; both lines of work begin from intelligence that another organization has already produced and shared.

Drawing on the practitioner survey, we divide this stage into four steps, illustrated in \autoref{fig:overview}: Intelligence Extraction, Normalization and Enrichment, Codification, and Distribution.
The four challenges the practitioner survey identified map onto these steps one to one; the first two steps generate CTI from local attack data, and the latter two prepare it for sharing.
While the steps resemble those of general knowledge engineering, each turns on a judgment specific to the domain; whether an artifact belongs to the attacker or to the victim, for example, cannot be determined from its form.
For each step, we describe its purpose, review the academic work that addresses it, and identify what still requires the analyst.

\subsection{Intelligence Extraction}
\label{ssec:filtration}

Intelligence extraction separates portable, actionable intelligence from the attack traces captured in logs.
As discussed in \autoref{sec:threat_data_collection}, these logs record confirmed attacks in detail, but the telemetry is tied to the victim organization's infrastructure and technology stack.
To produce CTI that other organizations can use, analysts must identify the artifacts that represent the attacker's capabilities and exclude those that merely record the victim's environment; in the scenario of \autoref{sec:lifecycle}, the address of the C\&C server is intelligence, while the hostnames of the compromised machines are not.

Prior work on extracting IoCs differs in the input it starts from.
\emph{CTI Consumption} systems extract indicators from published threat reports and security articles~\cite{liao2016acing, zhu2018chainsmith, mischinger2024ioc}, documents whose authors have already selected the artifacts worth reporting.
Systems that start from raw evidence must perform that selection themselves.
EIGER~\cite{kurogome2019eiger} and the Dridex study~\cite{rudman2016dridex} derive IoCs from malware sandbox reports, and SHIELD~\cite{sun2025alerts} produces IoCs, ATT\&CK technique mappings, and attack narratives directly from host-based IDS alerts.
However, these systems cannot reliably distinguish artifacts the attacker controls from legitimate instances that appear in the same traces and often share the same syntax.
For example, they may extract the file hashes of system utilities such as PowerShell as IoCs, even though these are standard components that attackers misuse rather than artifacts they introduced.
In the practitioner survey, \surveyExtractionChallengeNum{} of \totalParticipants{} participants identified extracting intelligence from noisy attack logs as a challenge, and analysts consequently perform this step manually.
Our first experiment tests whether an LLM can make this distinction, separating attacker artifacts from benign system activity in raw sandbox reports, where the two are described in the same vocabulary (\autoref{sec:evaluation}).

\subsection{Normalization and Enrichment}
\label{ssec:normalization}

Once the intelligence has been extracted, the behavior it records must be expressed in vocabularies that recipients understand.
Analysts map observed behavior to MITRE ATT\&CK techniques~\cite{mitre_attack} and exploited weaknesses to identifiers such as the Common Weakness Enumeration (CWE)~\cite{cwe} and the Common Platform Enumeration (CPE)~\cite{cpe}; in the scenario of \autoref{sec:lifecycle}, the recorded PowerShell execution corresponds to technique T1059.001 and the Log4Shell exploitation to T1190.
Practitioners then enrich the mapping with context it does not yet carry.
Where infrastructure overlaps with earlier campaigns, they attribute the activity to a known threat actor, and they attach Common Vulnerabilities and Exposures (CVE)~\cite{cve} identifiers and Common Vulnerability Scoring System (CVSS)~\cite{cvss} scores that convey severity.
Enrichment turns a behavioral record into intelligence that a recipient can prioritize and act on.

The academic work on this step likewise differs in the input it starts from.
\emph{CTI Consumption} research maps published threat reports to ATT\&CK techniques~\cite{husari2017ttpdrill, legoy2020automated}, where the report already states the behavior in prose that is often close to the technique's own description.
Starting instead from raw evidence, DeepCAPA~\cite{vasan2024deepcapa}, SODA~\cite{sajid2021soda}, and MAMBA~\cite{huang2021open} map sandbox telemetry such as API call traces to techniques, and HOLMES~\cite{milajerdi2019holmes}, RapSheet~\cite{hassan2020tactical}, and WATSON~\cite{zeng2021watson} map behaviors detected in provenance data to the same taxonomy.
These systems work well when the attack manifests as behavioral patterns their underlying detectors recognize.
When the behavior is implicit in low-level events, extracting techniques requires interpretation that the systems do not perform, and the analyst must supply it.
In contrast to normalization, the literature survey found no systematic treatment of the enrichment step.
In the practitioner survey, \surveyTtpChallengeNum{} of \totalParticipants{} participants reported that correlating observed behaviors with established threat taxonomies is a recurring challenge.
Our second experiment tests whether an LLM can map raw evidence to ATT\&CK techniques without being told which techniques to look for (\autoref{sec:evaluation}).

\subsection{Codification}
\label{ssec:policy}

Once CTI is generated, it must be prepared for dissemination; this preparation is the codification step introduced in \autoref{sec:practitioner_survey}.
Codification selects, filters, and transforms the intelligence so that it remains actionable for the recipient while protecting the organization's sensitive information, and the extent of these changes depends on the recipient's trust level, regulatory requirements, and the communication medium.
Its central mechanism is redaction, in which analysts remove PII and OII, such as employee names, internal addresses, and hostnames that could reveal network topology or asset information; in the scenario of \autoref{sec:lifecycle}, the analyst removes the employee's email address and the internal hostnames while retaining the attacker's C\&C address and the exploited CVE, which recipients need to assess their own exposure.

Redaction may appear to be an instance of named entity recognition, a task the CTI literature supports with annotated corpora and trained extractors~\cite{satyapanich2020casie, lange2024annoctr}.
The two tasks answer different questions.
An entity recognizer assigns categories, finding person names, organizations, addresses, and indicators, whereas redaction must decide which side a value belongs to and whether policy permits sharing it.
A hostname that belongs to the victim must be removed, while one of the same form that belongs to the attacker is the intelligence the report exists to convey, and both carry the same category label.
Approaches to privacy in CTI sharing likewise avoid this decision rather than automate it.
Huff et al.~\cite{huff2024privacy} protect fields of structured STIX objects, where the schema marks which values are sensitive, and other work shares model parameters instead of data~\cite{fereidooni2022fedcri}, examines selective disclosure from a licensing perspective~\cite{Shim2022OnDL}, or validates intelligence after it has been shared~\cite{geras2024big}.

The redaction of unstructured CTI therefore remains with the analyst, and both directions of error are costly.
Removing too much renders the intelligence less useful, and removing too little exposes the organization and risks legal liability under regulations such as the General Data Protection Regulation (GDPR)~\cite{GDPR}.
Practitioners cited this challenge more often than any other, with \surveyRedactionChallengeNum{} of \totalParticipants{} participants identifying the exposure of sensitive information as their main concern when sharing.
Our third experiment tests whether an LLM can separate the victim's details from the attacker's when both appear in the same report, once with the victim's privacy as the first priority and once with the report's utility first (\autoref{sec:evaluation}).

\subsection{Distribution}
\label{ssec:format}

Distribution converts the codified intelligence into the formats the organization's sharing channels require.
Which channel an organization uses depends on the intended audience.
Unrestricted sharing distributes CTI to any consumer, openly as Open-Source Intelligence (OSINT) or through commercial feeds, while restricted sharing exchanges it within private groups of established trust, such as an Information Sharing and Analysis Center (ISAC), typically as structured STIX~\cite{STIX21} bundles; in the scenario of \autoref{sec:lifecycle}, the same incident yields a public threat report and a STIX bundle for the organization's trusted partners.
Exchanging intelligence through these channels builds a collective knowledge repository that individual organizations could not assemble alone~\cite{bouwman2022helping}.

In the practitioner survey, \surveySharingPlatformNum{} of \totalParticipants{} participants exchange CTI through threat intelligence sharing platforms such as MISP, yet \surveyFormatChallengeNum{} of \totalParticipants{} identified translating CTI into the different formats these channels require as an obstacle.
The academic literature has studied this exchange from several angles, including platform design~\cite{Wagner2016MISPTD}, quality metrics~\cite{li2019reading}, the effect of sharing communities~\cite{bouwman2022helping, Jin2024SharingCT}, licensing~\cite{Shim2022OnDL}, and platform usability~\cite{stojkovski2021s, geras2023sharing}, and all of it evaluates intelligence that is already shared.
\emph{CTI Consumption} systems that emit structured formats exist as well, but they start from published reports rather than from an organization's own findings~\cite{satvat2021extractor, cheng2025ctinexus}.
Converting internally held CTI into the schema a platform requires, while preserving the semantic relationships between its entities, therefore remains the analyst's task.
Our fourth experiment tests whether an LLM can write the findings established in the preceding steps into a STIX~2.1 bundle that conforms to the standard and can be parsed by a reference implementation, a check that prior conversion systems do not perform (\autoref{sec:evaluation}).

Across the four steps, the work that remains manual is the interpretation and restructuring of text, from raw telemetry to standard vocabularies to sharing formats.
Whether current LLMs can perform this work is the question the next section answers, with one experiment for each step.

\section{Exploring AI-Driven Workflows for CTI}
\label{sec:evaluation}

The practitioner survey identified four recurring challenges in \emph{CTI Generation and Sharing}, and the same participants identified these tasks as the leading candidates for automation (\autoref{sec:practitioner_survey}).
We therefore designed four experiments, one for each step defined in \autoref{sec:cti_generation_and_sharing} (\autoref{tab:evaluation_design}).
Experiment~1 (Intelligence Extraction) tests whether a model can identify, in a raw sandbox report, the indicators attributable to the malware rather than to the operating system.
Experiment~2 (Normalization and Enrichment) tests whether a model can name the ATT\&CK techniques the same evidence demonstrates, without a list of candidates.
Experiment~3 (Codification) tests whether a model can remove the details that identify the victim organization from an incident report while retaining the details that describe the attacker.
Experiment~4 (Distribution) tests whether a model can write an analyst's CTI findings into a STIX~2.1 file that conforms to the standard and can be parsed by a reference implementation.

Our evaluation is exploratory.
We use \evalModelsNum{} LLMs as probes and ask two questions.
The first is whether current LLMs, as a class, can support each of the four steps.
We do not rank model providers, and comparing individual models is not our object.
The second is whether an open-weight LLM matches its proprietary counterparts.
Such a model can run on hardware the organization controls, which bears directly on the cost and privacy concerns the practitioner survey recorded.

\mypar{Scope.}
The closest existing benchmarks and systems address different tasks than our experiments.
CTIBench~\cite{alam2024ctibench} and AnnoCTR~\cite{lange2024annoctr} evaluate the extraction of threat entities and techniques from published CTI reports, a \emph{CTI Consumption} task systematized by prior work~\cite{Buechel2025SoK}.
Our inputs are the raw evidence such reports are written from.
Systems such as DeepCAPA~\cite{vasan2024deepcapa} and SODA~\cite{sajid2021soda} map binaries and API call sequences to ATT\&CK techniques without reading reports.
As trained extractors, they are tied to one input format and to a fixed label space that must be relearned when ATT\&CK changes, whereas an LLM can receive the current taxonomy in context without retraining.
For redaction, existing CTI privacy approaches operate on structured STIX fields~\cite{huff2024privacy} and do not support configurable policies over report text.
For distribution, published systems that emit STIX report how accurately they recover the entities and relationships a report describes, but they do not check whether the resulting file is accepted by the software meant to read it, and a file rejected on receipt conveys nothing.

\begin{table}[t]
\centering
\caption{Experiment design overview.}
\label{tab:evaluation_design}
\scriptsize
\setlength{\tabcolsep}{2pt}
\renewcommand{\arraystretch}{1.15}
\arrayrulecolor{black!35}
\begin{tabular}{|>{\raggedright\arraybackslash}p{0.24\columnwidth}|>{\raggedright\arraybackslash}p{0.34\columnwidth}|>{\raggedright\arraybackslash}p{0.34\columnwidth}|}
\hline
\rowcolor{lightgray}
\textbf{Framework Step} & \textbf{Input} & \textbf{Output} \\
\hline
Intelligence Extraction & Sandbox report & IoCs (files, registry keys, mutexes, network) \\
\rowcolor{lightergray}
Normalization \& Enrichment & Sandbox report, campaign network capture & ATT\&CK techniques \\
Codification & Incident report with an injected victim identity & Values to redact, with replacements \\
\rowcolor{lightergray}
Distribution & CTI findings (software, techniques, indicators, steps) & STIX~2.1 file \\
\hline
\end{tabular}
\arrayrulecolor{black}
\renewcommand{\arraystretch}{1.0}
\end{table}

\subsection{Datasets and Labels}
\label{ssec:eval_datasets}

Published threat reports are finished analyses; the evidence behind them is almost never released.
The raw data also carries no labels.
A sandbox report lists thousands of file, registry, and network events, but it does not mark the events that matter.
Separating attacker activity from ordinary system behavior is the judgment these experiments measure; the labels therefore cannot be inherited from the data.
We built both the evidence and the labels.

\mypar{Malware evidence.}
We collected \evalSamplesNum{} malware samples spanning different families and executed each one in a security vendor's instrumented sandbox, which produced one report per sample.

\mypar{Campaign evidence.}
For intrusions that unfold across multiple hosts and steps, we replayed \evalCampaignsNum{} real campaigns: BlackSuit~\cite{dfir_blacksuit_2024}, Confluence LockBit~\cite{dfir_lockbit_2025}, Egg-Cellent Resume~\cite{dfir_resume_2024}, Lynx~\cite{dfir_lynx_2025}, and Nitrogen~\cite{dfir_nitrogen_2024}.
The replays ran in a controlled network we built to resemble a small enterprise, with a domain controller, workstations, a gateway, and a server operated by the attacker.
We reconstructed each campaign from the full text of its published incident report, in every case an article from The DFIR Report~\cite{dfir}, and executed the campaign step by step.
Each campaign comprised between 10 and 22 steps across 4 to 8 hosts.
We captured network traffic at every host as each step ran.
The evidence therefore records each attack as it happened rather than describing it afterward.

\mypar{Redaction inputs.}
The inputs are the full text of the five DFIR Report articles behind the campaigns; these reports cannot be redacted as they stand, because their authors removed the victim's identity before publication.
For each of the five reports, which average around 5{,}400 words, we injected a victim organization's identity into the text.
Some injected values identify a person or an asset directly, such as names, email addresses, employee numbers, and internal network addresses.
Others are derived from those names, such as machine names, internal Windows domains, and service accounts, so that the link back to the original identity is real.
Each value was placed where a value of its kind naturally appears in the report, for example an email address inside the output of a tool that harvests credentials.
The experiment therefore measures redaction rather than the detection of fabricated facts.
In total we injected \evalPlantedIdentifiersNum{} identifying values across the five reports.
We score \evalScoredIdentifiersNum{} distinct values, of which \evalHeldoutIdentifiersNum{} appear in the four reports reserved for the final evaluation.
In addition, we injected \evalDescriptivePhrasesNum{} descriptive phrases that give the injected organization's industry, size, and region.
Such a phrase matches no pattern, but together the phrases narrow the organization to a small set of candidates, and a model may remove a phrase entirely or in part; we therefore score them separately from the identifying values.

\mypar{Distribution inputs.}
Experiment~4 does not start from raw evidence, but from a CTI findings report for each campaign.
The model is tasked with three judgments that these findings leave open: whether each software is an ordinary tool the intruder misused or malware they brought, which technique each software was used for, and which recovered file hash identifies which software.

\mypar{Labels.}
We produced initial labels from the annotations the sandbox vendor generates automatically for the malware reports, and from the published incident reports together with our own execution records for the campaigns.
For each experiment, two analysts then independently developed the final labels from this information and the underlying evidence, and resolved every disagreement through discussion; every label therefore reflects inter-analyst agreement.

\subsection{Models, Protocol, and Baselines}
\label{ssec:eval_setup}

\mypar{Models.}
We evaluate three recent general-purpose LLMs, \modelPropA{}, \modelPropB{}, and the open-weight \modelOpen{}.
All three models were accessed through LiteLLM~\cite{litellm} with the temperature set to 0.7.
Below we refer to them as the two proprietary models and the open-weight model, and we report per-model numbers where the models diverge.

\mypar{Input size.}
Because LLMs have a limited context window, we retained only the sandbox reports smaller than 700\,KB.

\mypar{Tuning and runs.}
For each experiment we reserved a tuning set of \evalTuningReportsNum{} sandbox reports and one campaign, BlackSuit, and used it to tune the prompts, the output formats, and the scoring code through three prompt versions.
The remaining \evalHeldoutReportsNum{} reports and \evalHeldoutCampaignsNum{} campaigns were held out as the testing set and scored once with the final prompts.
Each held-out input was run three times per model, and an item is reported only if it appears in at least two of the three runs.

\mypar{Baselines.}
For Experiments~1 and~2 we compare against rule-based baselines.
For Experiment~1, we use two static rules: a permissive rule that captures every artifact in the report, and a restrictive rule that captures only the artifacts the sample created or modified.
These rules reflect standard sandbox IoC extractors.
Every labeled indicator is present in the report; the first rule therefore cannot miss one, and the comparison isolates the judgment of what to include.
For Experiment~2 we use Sigma~\cite{sigmahq}, an open library of detection rules maintained by the security community, in which each rule is annotated with the ATT\&CK techniques it detects.
We chose Sigma because it is independent of our data, it is maintained against current ATT\&CK, and rules exist for 61 of the 62 techniques in our labels.
Sigma expects Windows event logs; we therefore wrote a translation layer that renders each sandbox report, noise included, in the event form its rules expect.
Both the rules and the models are scored under two conditions, one that counts a mapping as correct when the main technique matches and a stricter one that requires the sub-technique to match as well.

Experiments~3 and~4 have no baseline, for different reasons.
For redaction, existing tools recognize categories of values, such as names and addresses, but the task is to decide whether a value belongs to the victim or to the attacker.
Existing tools do not attempt this ownership decision, and we treat that absence as part of the result (\autoref{sec:results}).
For distribution, prior systems convert threat reports rather than CTI findings and do not check acceptance; a comparison would therefore compare two different tasks.

\subsection{Results}
\label{sec:results}

\begin{table}[t]
\centering
\caption{Experiments~1 and~2 over the \evalHeldoutReportsNum{} held-out sandbox reports.
(a) Indicator extraction, 1,472 labeled indicators (Total).
(b) ATT\&CK technique mapping, 254 labeled main techniques and 263 labeled sub-techniques; the Sigma rules do not identify sub-techniques.}
\label{tab:eval_generation}
\scriptsize
\setlength{\tabcolsep}{3pt}
\renewcommand{\arraystretch}{1.05}
\arrayrulecolor{black!35}
\textbf{(a) Indicator extraction}\par\smallskip
\resizebox{\columnwidth}{!}{%
\begin{tabular}{|l|c|c|c|c|c|}
\hline
\rowcolor{lightgray}
\textbf{Approach} & \textbf{Total} & \textbf{TP} & \textbf{FP} & \textbf{FN} & \textbf{Prec.} \\
\hline
Rule: every recorded artifact & 1,472 & 1,472 & 15,019 & 0 & 0.09 \\
\rowcolor{lightergray}
Rule: created or modified & 1,472 & 1,472 & 2,115 & 0 & 0.41 \\
\modelPropA{} & 1,472 & 365 & 105 & 1,107 & 0.78 \\
\rowcolor{lightergray}
\modelPropB{} & 1,472 & 363 & 145 & 1,109 & 0.71 \\
\modelOpen{} & 1,472 & 461 & 451 & 1,011 & 0.51 \\
\hline
\end{tabular}}
\par\smallskip
\textbf{(b) ATT\&CK technique mapping}\par\smallskip
\resizebox{\columnwidth}{!}{%
\begin{tabular}{|l|c|c|c|c|c|c|c|c|}
\hline
\rowcolor{lightgray}
 & \multicolumn{4}{c|}{\textbf{Technique}} & \multicolumn{4}{c|}{\textbf{Sub-technique}} \\
\rowcolor{lightgray}
\textbf{Approach} & \textbf{TP} & \textbf{FP} & \textbf{FN} & \textbf{F1} & \textbf{TP} & \textbf{FP} & \textbf{FN} & \textbf{F1} \\
\hline
Sigma rules & 37 & 160 & 217 & 0.16 & $\times$ & $\times$ & $\times$ & $\times$ \\
\rowcolor{lightergray}
\modelPropA{} & 227 & 161 & 27 & 0.71 & 224 & 183 & 39 & 0.67 \\
\modelPropB{} & 201 & 102 & 53 & 0.72 & 193 & 124 & 70 & 0.67 \\
\rowcolor{lightergray}
\modelOpen{} & 186 & 74 & 68 & 0.72 & 176 & 92 & 87 & 0.66 \\
\hline
\end{tabular}}
\arrayrulecolor{black}
\renewcommand{\arraystretch}{1.0}
\end{table}

\mypar{Experiments 1 and 2 (Intelligence Extraction; Normalization and Enrichment).}
In the practitioner survey, \surveyExtractionChallengeNum{} of \totalParticipants{} participants reported extracting indicators from noisy attack data as a recurring challenge, and \surveyTtpChallengeNum{} of \totalParticipants{} reported correlating observed behaviors with standardized TTPs as another recurring challenge.
The first two experiments test these two steps on the same evidence, and the models share the same limitation on both; we therefore present them together.

In Experiment~1, the three LLMs and the rule-based baselines are given a raw sandbox execution log and are tasked with identifying the malicious indicators among the files, registry keys, mutexes, and network artifacts it records.
\autoref{tab:eval_generation}a provides the results.
For indicators, we report precision because the two directions of error are not equally costly.
Shared IoCs are consumed as blocklists.
A benign value reported as an indicator raises false detections in every organization that blocks it, whereas a missed indicator costs one detection opportunity.
As expected, the rule-based baselines recover every labeled indicator, but they over-report.
The permissive rule reports 15,019 artifacts that are not indicators, for a precision of 0.09, and the restrictive rule 2,115, for a precision of 0.41.
A static rule cannot distinguish an artifact the malware introduced from one that ordinary system activity produces.
The three LLMs are more conservative.
They report far fewer values than the rules and reach a precision between 0.51 and 0.78.
The difference is largest for file indicators, where the two rules reach precisions of 0.07 and 0.40 while the three models average 0.80.
However, reporting so few values costs completeness.
The models recover only 363 to 461 of the 1,472 labeled indicators, a recall of 0.25 to 0.31.
A report carries roughly 33 labeled indicators on average, many of them nearly identical paths produced by one behavior, and a model returns roughly 10.
The models rely on how a value looks.
They report values that appear malicious and overlook mundane ones, such as paths under legitimate software directories that the sample nevertheless modified.
Recovering such values requires tracing them back to the behavior that produced them, and a single pass over a report with thousands of events cannot do that.

In Experiment~2, the models receive the same sandbox reports as in Experiment~1 and must identify the ATT\&CK techniques the evidence demonstrates.
The baseline is Sigma, a library of static detection rules written by the security community (\autoref{ssec:eval_setup}).
Because the Sigma rules expect Windows event logs, we run them on the sandbox reports rendered in that form.
\autoref{tab:eval_generation}b provides the results.
The three models recover most of the labeled techniques, with an F1 of 0.71 to 0.72 against 0.16 for Sigma.
Sigma performs poorly because its rules look for specific event patterns.
Almost half of the applicable rules look for suspicious process launches and command lines, and a malware sample that works through direct system calls produces few of either.
The models, in contrast, are format agnostic.
Under the second condition, a prediction counts as correct only when the sub-technique also matches.
Sigma does not identify sub-techniques.
The models, drawing on their broad knowledge of ATT\&CK, lose almost nothing under this condition; the largest drop is six points of F1.
When a model names the correct parent technique, it usually also names the correct sub-technique.
The models' limitation mirrors Experiment~1.
They assert techniques whose evidence the report does not contain and miss techniques whose evidence is plainly recorded; masquerading (T1036) is labeled in 21 reports and missed by all three models.
For ten techniques, every model both misses them in reports where they are labeled and asserts them in reports where they are not.
The difficulty is therefore not knowing the techniques but deciding whether the evidence in a particular report warrants them.

\evalinsight{LLMs separate malicious indicators from benign noise and recover techniques that static rules cannot, with far fewer false positives.
However, they recover only a fraction of the labeled indicators and assert techniques the evidence does not support; addressing both calls for task decomposition and multiple passes over the input.}

\begin{table}[t]
\centering
\caption{Experiment~3, redaction over the \evalHeldoutCampaignsNum{} held-out reports with \evalHeldoutIdentifiersNum{} injected values (Total) and \evalDescriptivePhrasesNum{} descriptive phrases.}
\label{tab:eval_redaction}
\scriptsize
\setlength{\tabcolsep}{3pt}
\renewcommand{\arraystretch}{1.05}
\arrayrulecolor{black!35}
\resizebox{\columnwidth}{!}{%
\begin{tabular}{|l|l|c|c|c|c|c|c|}
\hline
\rowcolor{lightgray}
\textbf{Model} & \textbf{Prompt} & \textbf{Total} & \textbf{TP} & \textbf{FP} & \textbf{FN} & \textbf{F1} & \textbf{Phrases removed} \\
\hline
\modelPropA{} & privacy preserving & 92 & 92 & 1 & 0 & 0.99 & 10 of 10 \\
\rowcolor{lightergray}
\modelPropB{} & privacy preserving & 92 & 91 & 50 & 1 & 0.78 & 10 of 10 \\
\modelOpen{} & privacy preserving & 92 & 89 & 11 & 3 & 0.93 & 7 of 10 \\
\hline
\rowcolor{lightergray}
\modelPropA{} & utility preserving & 92 & 84 & 12 & 8 & 0.89 & 5 of 10 \\
\modelPropB{} & utility preserving & 92 & 82 & 21 & 10 & 0.84 & 3 of 10 \\
\rowcolor{lightergray}
\modelOpen{} & utility preserving & 92 & 89 & 12 & 3 & 0.92 & 1 of 10 \\
\hline
\end{tabular}}
\arrayrulecolor{black}
\renewcommand{\arraystretch}{1.0}
\end{table}

\mypar{Experiment 3 (Codification).}
In the practitioner survey, \surveyRedactionChallengeNum{} of \totalParticipants{} participants reported difficulty preventing the exposure of sensitive information, the most widespread of the four challenges.
The experiment gives a model one incident report injected with PII and OII.
The model must return the values that should be removed, each with a replacement.
The core challenge in this task is ownership.
A report contains machine names, addresses, and accounts of both the victim and the attacker, written in the same form, and the form of a value does not reveal which side it belongs to.
This is also why the experiment has no baseline.
Existing redaction tools recognize categories of values; a rule that removes every machine name removes the attacker's along with the victim's and renders the resulting report far less useful.

We run each report under two prompts, one that tells the model to preserve the victim's privacy and one that tells it to preserve the report's utility.
Under the privacy-preserving prompt, the goal is that no reader can determine which organization was attacked.
Under the utility-preserving prompt, the goal is that the intelligence arrives intact.
\autoref{tab:eval_redaction} provides the results.
The models adhere strictly to the overall goal of their prompt, and the two prompts therefore produce opposite errors.
Under the privacy-preserving prompt, the models redact almost all of the injected PII and OII, missing at most 3 of the \evalHeldoutIdentifiersNum{} injected values, and two of the three remove all \evalDescriptivePhrasesNum{} descriptive phrases.
However, they also remove content that identifies no one.
\modelPropB{} removes 50 values that were never identifiers, 35 percent of everything it removes.
Under the utility-preserving prompt, the errors reverse.
\modelPropA{} now misses 8 of the 92 injected values and \modelPropB{} misses 10, and the models keep more of the descriptive phrases; \modelPropA{} removes 5 of the 10, \modelPropB{} 3, and \modelOpen{} only 1.
The two prompts also reverse the treatment of individual values.
All three models miss the same three injected internal addresses under the utility-preserving prompt; under the privacy-preserving prompt, every injected value is flagged by at least one model.

This trade-off is not an artifact of our injected identities.
An internal address can be victim infrastructure and useful intelligence at the same time.
When the address marks the attacker's route through the network, removing it removes the observation the report was written to share.
Whether such a detail remains or is removed is a policy question about what keeps CTI useful.
Existing redaction work does not study this trade-off and offers no such policies.
Without such a policy, the models over-redact when told to preserve privacy and leak when told to preserve utility.

\evalinsight{The models redact according to whichever goal their prompt states; privacy first produces over-redaction, and utility first produces leaks.
Balancing the two requires redaction policies that state which details a shared report must keep and which it may lose; future work should develop such policies.}

\begin{table}[t]
\centering
\caption{Experiment~4 over the \evalHeldoutCampaignsNum{} held-out intrusions: (a) the software type and (b) the links.}
\label{tab:eval_stix}
\scriptsize
\setlength{\tabcolsep}{3pt}
\renewcommand{\arraystretch}{1.05}
\arrayrulecolor{black!35}
\textbf{(a) Software type}\par\smallskip
\resizebox{0.9\columnwidth}{!}{%
\begin{tabular}{|l|c|c|c|c|c|}
\hline
\rowcolor{lightgray}
\textbf{Model} & \textbf{Total} & \textbf{TP} & \textbf{FP} & \textbf{FN} & \textbf{F1} \\
\hline
\modelPropA{} & 21 & 19 & 2 & 2 & 0.90 \\
\rowcolor{lightergray}
\modelPropB{} & 21 & 19 & 2 & 2 & 0.90 \\
\modelOpen{} & 21 & 19 & 2 & 2 & 0.90 \\
\hline
\end{tabular}}
\par\smallskip
\textbf{(b) Links}\par\smallskip
\resizebox{0.9\columnwidth}{!}{%
\begin{tabular}{|l|c|c|c|c|c|}
\hline
\rowcolor{lightgray}
\textbf{Model} & \textbf{Total} & \textbf{TP} & \textbf{FP} & \textbf{FN} & \textbf{F1} \\
\hline
\modelPropA{} & 38 & 34 & 14 & 4 & 0.79 \\
\rowcolor{lightergray}
\modelPropB{} & 38 & 38 & 29 & 0 & 0.72 \\
\modelOpen{} & 38 & 38 & 35 & 0 & 0.68 \\
\hline
\end{tabular}}
\arrayrulecolor{black}
\renewcommand{\arraystretch}{1.0}
\end{table}

\mypar{Experiment 4 (Distribution).}
In the practitioner survey, \surveyFormatChallengeNum{} of \totalParticipants{} participants reported that translating CTI into the formats sharing platforms require is a recurring obstacle.
In this experiment, a model receives the CTI findings for one intrusion, which list the software, the techniques, the indicators, and the steps of the attack, and must write them into a single STIX~2.1 file.
Every model was able to produce a file in the STIX format that the \texttt{stix2} reference library parses.
Every file carries all of the supplied techniques and indicators.
We therefore evaluate the judgments the findings leave open: the software type and the two kinds of links.

The software type states whether a software is a \texttt{tool} or \texttt{malware}.
In STIX, a \texttt{tool} is legitimate software that an intruder used, and \texttt{malware} is software written to do harm.
The type decides how a recipient responds; malware can be blocked immediately, whereas a tool may also be running legitimately in the recipient's own network.
The links connect the objects of the file.
One kind of link states which technique a software was used for, and the other states which recovered file hash identifies which software.
The links carry the story of the intrusion; without them, the file is a set of disconnected lists.
A link counts as correct only when it connects the right pair of objects with the right relationship type.
\autoref{tab:eval_stix} provides the results.

On the software type, the three models perform identically and correctly type 19 of the 21 software.
The models identify the software reliably, including from the file names behind the recovered hashes.
On the links, every model asserts more than the findings support.
\modelOpen{} asserts 73 links where 38 are correct, and \modelPropB{} asserts 67.
For example, a hash recovered for one software is also linked to other software, and a software is linked to techniques it was not used for.
A recipient therefore cannot trust an individual link without checking it.
Future work should validate what a generated file claims and not only its format, for example by requiring every link to name the finding it rests on.

\evalinsight{The models produce valid STIX files and identify the software type reliably.
However, every model asserts links the findings do not support; future work should validate the links a shared file asserts and not only its format.}

\mypar{The open-weight model, cost, and confidentiality.}
Two of the hurdles the practitioner survey recorded are the cost of AI (\surveyAiCostHurdleNum{} of \totalParticipants{} participants) and data privacy (\surveyAiPrivacyHurdleNum{} of \totalParticipants{}).
A model whose weights are published can address both of these, and we evaluate \modelOpen{} as one such model.
At list prices, running the reserved test data for all four experiments costs approximately \evalCostTestPropB{} for \modelPropB{}, \evalCostTestPropA{} for \modelPropA{}, and \evalCostTestOpen{} for \modelOpen{}.
\modelOpen{} matches the proprietary models at technique mapping and at redaction under both prompts.
Its weaknesses are indicator precision, where 0.51 means an analyst reviews roughly one wrong indicator for every right one, and links, where it asserts the most unsupported relationships.
On confidentiality, the evidence these experiments process is precisely the information an organization is reluctant to share.
Sending an unredacted incident report to a third party for redaction means disclosing exactly what the organization is trying to protect.
A model running on hardware the organization controls removes that contradiction, because the report never leaves the organization, and for redaction this is closer to a requirement than a preference.
A model of this capability that an organization can host itself therefore makes it practical to build the workflows of several steps that the errors above call for, without paying per token for every step.

\section{Discussion and Future Research Directions}
\label{sec:discussion}

The four experiments in \autoref{sec:evaluation} measure where current models support \emph{CTI Generation and Sharing} and where they fall short, and each shortfall points to a research direction.

\mypar{Task Decomposition and Agentic Workflows.}
The two generation steps, Intelligence Extraction and Normalization and Enrichment, leave the most room for improvement.
On Intelligence Extraction the models are conservative, recovering a quarter to a third of the labeled indicators across the held-out reports.
On Normalization and Enrichment, roughly a third of the technique mappings a report needs are wrong or absent (F1 between 0.71 and 0.72).
Both experiments give the model an entire report and a single instruction; it must therefore separate attacker activity from system noise, recognize the behavior, and name it in a single pass.
We expect agentic workflows, in which the model works through the evidence as a sequence of smaller tasks rather than in a single pass, to improve both steps.
A sandbox report can be divided, by process or by stage of the attack, into segments small enough to examine closely.
Candidate indicators and techniques can first be proposed and then separately verified.
Every accepted claim can carry the events that support it, so that an analyst reviews that evidence instead of rereading the report.
Decomposition also becomes necessary when the evidence exceeds the model's context window.
The largest reports in our evaluation already approach 540\,KB, and larger incidents can exceed any model's window.

Such a workflow should also validate its own output.
Our distribution results show that the models produce files in the correct format but assert links the findings do not support, and a wrong link passes every format check.
A validator that checks what a file claims, and whose verdict is fed back to the model, would let the workflow correct a file before it is sent rather than leaving the defect for the recipient to discover.

\mypar{Datasets for CTI Generation and Sharing.}
Measuring progress on these steps requires labeled data, and obtaining it was our main obstacle; as \autoref{ssec:eval_datasets} describes, we had to build both the evidence and the labels ourselves, with two analysts labeling independently and resolving their disagreements through discussion.
Future benchmarks should preserve every step from raw evidence to shareable intelligence for the same incident, so that errors can be traced across the pipeline instead of measured one step at a time.

The scope of enrichment should also grow beyond what we evaluated.
We used ATT\&CK techniques as the enrichment vocabulary, but practitioners add more than techniques.
Extending the evaluation to CVE identifiers~\cite{cve}, attack patterns from the Common Attack Pattern Enumeration and Classification (CAPEC)~\cite{capec}, threat actor attribution, and severity context would make the resulting CTI more actionable for its recipients.

\mypar{Redaction Policies.}
The redaction results show that Codification cannot be treated as an entity recognition task, and that future work should supply the policy rather than a better model.
The models follow whichever goal their prompt states.
However, they cannot weigh a small gain in usefulness against a small loss of anonymity, because that trade-off is a policy decision their input does not contain.
Errors in both directions are costly, and the scoring hides how costly each one is; removing a boilerplate phrase and removing the attacker's server address count as the same false positive.
Future work should study how redaction policies are constructed and applied: what threshold of identifiability a policy should set, what removals cost readers, how redacted values can be replaced so that the report remains usable, and how strict and permissive policies change the intelligence that reaches recipients. Over-redaction in particular should be measured by its consequence rather than by its count.

\section{Conclusion}

While CTI is essential for defending organizations, our systematization reveals a gap between research and practice.
The literature survey of \filteredPapersNum{} papers shows that \emph{Threat Data Collection} and \emph{CTI Consumption} are well studied, whereas \emph{CTI Generation and Sharing} is treated as a procedural task rather than a research problem.
The practitioner survey confirms that this stage is largely manual, with redaction, extraction, and standardization as the primary bottlenecks.

Our evaluation shows that the models report far fewer false indicators than the rule-based baselines and recover most of the labeled ATT\&CK techniques where static rules recover almost none.
However, they recover only a quarter to a third of the labeled indicators.
During PII and OII redaction, the models follow the priority they are given but do not judge what keeps the report useful.
The STIX files they produce carry every finding yet assert relationships the findings do not support.
These errors recur across all three evaluated models and stem from how the tasks are posed rather than from any one model.

Addressing these errors requires research specific to this domain: workflows that decompose the task and tie each claim to its evidence, redaction guided by explicit sharing policies, validators that check what a file claims and not only its format, and benchmarks that span every step from raw evidence to shared intelligence.
The open-weight model we evaluate is competitive at technique mapping and redaction at a lower cost; this research can therefore build on models that run on hardware the organization controls.
Our framework, datasets, and experimental methodology, which we will share with the community, provide a foundation for this work.

\section{Acknowledgement}
We would like to thank Anton Malyshenok for their help with the dataset used in our evaluation.
This material is partially based on work supported by the Office of Naval Research under award number N00014-23-1-2387 and by the National Science Foundation under grant no. 2229876, and is also supported in part by funds provided by the National Science Foundation, by the Department of Homeland Security, and by IBM.
The opinions, findings, conclusions, or recommendations expressed in this material are those of the author(s) and do not necessarily reflect the views of the ONR, NSF, or any of their federal agency or industry partners.

\clearpage

\section*{Ethics Considerations}

We present a stakeholder-based ethical analysis of our work, guided by the Menlo Report principles of Beneficence, Respect for Persons, Justice, and Respect for Law and Public Interest, and by USENIX Security's ethics guidelines.
We consider both the ethics of our research process and the potential impacts of publication.

\mypar{Stakeholders}
Our work involves the following stakeholder groups:
\begin{itemize}
    \item practitioner participants who contributed to the practitioner survey;
    \item organizations that generate and share CTI, whose workflows and constraints are reflected in our analysis;
    \item CTI consumers, including defenders who rely on shared intelligence;
    \item the research community, which may build on our framework and findings; and
    \item potential adversaries, given the dual-use nature of CTI automation techniques.
\end{itemize}

\mypar{Research Process Impacts}
\textit{\mypar{Respect for persons and privacy.}}
Our practitioner study involved \totalParticipants{} analysts across \totalOrganizations{} organizations.
Participation was voluntary and involved no deception.
We collected only high-level workflow descriptions, excluding raw telemetry, proprietary CTI, or sensitive operational data.

Participants received a questionnaire of \numberofquestions multiple-choice questions.
Respondents could select one or more answers for each question.
We also included an ``Other'' option for open-ended textual responses.
The complete questionnaire and aggregated responses appear in \autoref{tab:survey_questions}.
Results are reported in aggregate to prevent re-identification of individuals or organizations.
\textit{\mypar{Evidence production.}}
The evidence used in our LLM evaluation was produced by us.
We executed \evalSamplesNum{} malware samples in the instrumented sandbox described in \autoref{sec:evaluation}, and we replayed \evalCampaignsNum{} campaigns, reconstructed from published incident reports~\cite{dfir}, in a controlled network that we built for this purpose.
Every system attacked during the replays, including the server that acted as the attacker's infrastructure, was our own, and no production system was involved.

\textit{\mypar{Malware handling.}}
We executed the malware samples only by submitting them to the instrumented sandbox, and we developed no new malware, vulnerabilities, or exploits for this work.
The released artifacts identify each sample by its cryptographic hash rather than redistributing the sample itself.

\textit{\mypar{Human subjects and IRB.}}
The practitioner survey did not require IRB review because it did not collect or retain identifiable private information about the respondents.
We did not record participant names, email addresses, job titles, employer names, IP addresses, or any other attribute that could characterize or re-identify an individual respondent.
The questionnaire elicited only closed-form and free-text descriptions of professional CTI workflows, and every response was stored and analyzed in aggregate form from the outset.
Participation was voluntary, participants were informed of the purpose and scope of the study before responding, and they could decline any question or withdraw at any time.

\mypar{Publication Impacts and Risks}
\textit{\mypar{Beneficence.}}
Our work aims to improve defensive cybersecurity practice by systematizing the underexplored \emph{CTI Generation and Sharing} phase and by empirically assessing where LLMs can and cannot meaningfully assist analysts.
\textit{\mypar{Dual-use considerations.}}
Automation techniques for CTI generation may be misused by adversaries to scale analysis or evade detection.
However, our work does not introduce new vulnerabilities, exploits, or attack techniques.
We mitigate the risk of misuse by emphasizing analyst-in-the-loop workflows, documenting systematic model errors, and avoiding claims of end-to-end autonomous CTI generation.
\textit{\mypar{Balancing privacy and utility.}}
Our redaction experiment shows that the models respond to the priority stated in their instructions, removing more of the victim's descriptive context under the privacy-preserving prompt and less under the utility-preserving one.
However, they cannot weigh what a detail reveals about the victim against its value to the recipient, because the input contains no policy for this decision.
We therefore frame redaction performed by a language model as assistive rather than authoritative.

\mypar{Use of AI Assistants}
All text and figures in this paper were created by the authors.
Where LLMs were used during writing, their role was limited to grammar-suggestion and style-suggestion tools, and all suggestions were manually reviewed and accepted by an author. LLMs were also used to assist in implementing parts of the experimental workflow; however, any generated code underwent the same validation and testing procedures as code written by a human developer.

\textit{\mypar{Reproducibility caveat.}}
Proprietary LLMs evolve, which may affect exact reproducibility of the experiments in \autoref{sec:evaluation}.
As the underlying models advance, both the results and the errors may change; we therefore name the evaluated models and their versions in \autoref{sec:evaluation}, record the configurations we logged in the released artifacts, and encourage future work to re-evaluate these trends.

\mypar{Law, Public Interest, and Mitigations}
Our research complies with applicable laws and institutional policies.
We did not conduct experiments on live systems without consent.
We did not violate the terms of service of the platforms and data sources we used, and we discovered no vulnerabilities that required disclosure.
Mitigations include limiting the sensitivity of the data we collected, aggregating survey responses, and documenting risks and limitations.
Residual dual-use risks remain and are described under Dual-use considerations above.

\mypar{Decision to Conduct and Publish}
We decided to conduct and publish this research after weighing ethical benefits against potential harms.
From a beneficence perspective, improving the rigor and safety of CTI workflows outweighs foreseeable risks given our mitigations.
From a rights-based perspective, we avoided violations of privacy, consent, and confidentiality.
We conclude that transparent publication, coupled with explicit discussion of limitations and safeguards, best serves the public interest and the cybersecurity community.

\clearpage

\section*{Open Science}

We are committed to the USENIX Security open science policy.
This SoK is self-contained, in that the artifacts needed to evaluate our contributions are either reported in the paper or released through the anonymous repository below.

\mypar{Artifacts Contained in the Paper}
Because this is a systematization paper, two of our artifacts are reported directly in the manuscript.
\begin{itemize}[leftmargin=1.2em,itemsep=1pt,topsep=2pt]
  \item \textit{Literature corpus.} The \filteredPapersNum{} papers that form our systematization are shown, together with the stage we assign to each, in \autoref{fig:cti_mindmap}.
The venues covered and the exact Google Scholar and DBLP queries used to build the corpus are listed in Appendix~\ref{appendix:paper_selection}; the search can therefore be reproduced independently.
  \item \textit{Practitioner survey.} The full questionnaire and the aggregated responses of the \totalParticipants{} participants are provided in \autoref{tab:survey_questions} in Appendix~\ref{appendix:survey_questionnaire}.
\end{itemize}

\mypar{Artifacts Released Anonymously}
The remaining artifacts are available to the reviewers through an anonymous, tracking-free repository at \artifacturl
In this preprint, the repository identifier is masked, because the link is reserved for the reviewers of the submitted version.
The repository contains the prompts of the four experiments, the datasets and their labels, the model outputs, the scripts that run the experiments and produce the final results reported in the paper, and the classification of every surveyed paper with the reasoning for its placement.
The surveyed papers themselves are not redistributed, because redistribution is subject to the respective publishers' terms; each paper is identified by its title, from which the published version can be located through a bibliographic search.
The captures and step records of the replayed campaigns describe attacks that are already documented in the published incident reports they were reconstructed from; their release therefore adds no new attack capability.

\mypar{Limitations on Release}
We do not retain individual responses to the practitioner survey.
As described in the Ethics Considerations section, responses were collected and stored only in aggregate form, because data at finer granularity could enable re-identification of participants or of their organizations.
The aggregated results in \autoref{tab:survey_questions} are the complete survey data we hold.
The malware samples themselves are not redistributed; each is identified in the released reports and labels by its hash.
We will make all releasable artifacts publicly available under a permanent, non-anonymous link upon acceptance of the submitted version.

\clearpage

\bibliographystyle{IEEEtran}
\bibliography{IEEEabrv, refs}

\clearpage

\label{sec:appendix}
\appendix
\section{Paper Selection}
\label{appendix:paper_selection}

\mypar{Conferences Covered}

\begin{appendixbox}
ACSAC, ACSW, AsiaCCS, IEEE BigData, BRAINS, CCS, CyCon, DBSec, DIMVA, DSN,
ESORICS, EuroS\&P, EuroS\&P Workshops, ICDE, ICISS, IEEE S\&P, IJCNN, IMC, ISI, ISPEC,
NDSS, RAID, SIGMOD, USENIX Security, WISCS
\end{appendixbox}

\mypar{Google Scholar Queries}

\begin{appendixbox}
``cyber threat intelligence'' extraction OR mining OR classification,
``TTP'' extraction OR mining OR classification,
``att\&ck'' extraction OR mining OR classification,
``threat report'' extraction OR mining OR classification,
``attack pattern'' extraction OR mining OR classification,
``attack technique'' extraction OR mining OR classification,
``attack graph'' extraction OR mining OR classification,
``threat sharing'' OR ``threat intelligence sharing'',
``threat report generation'' OR ``CTI generation'',
``threat mining'' OR ``cyber threat mining'',
``threat investigation'' OR ``cyber threat investigation''
\end{appendixbox}

\mypar{DBLP Queries}

\begin{appendixbox}
``cyber threat intelligence'' extraction|mining|classification,
``TTP'' extraction|mining|classification,
``att\&ck'' extraction|mining|classification,
``threat report'' extraction|mining|classification,
``attack pattern'' extraction|mining|classification,
``attack technique'' extraction|mining|classification,
``attack graph'' extraction|mining|classification,
``threat sharing''|``threat intelligence sharing'',
``threat report generation''|``CTI generation'',
``threat mining''|``cyber threat mining'',
``threat investigation''|``cyber threat investigation'',
``threat intelligence'',
``threat sharing'',
``TTP'',
``threat detection'',
``threat analysis'',
``cyber threat intelligence'',
``threat report'',
``threat investigation''
\end{appendixbox}

\clearpage
\onecolumn
\section{Survey Questionnaire and Results}
\label{appendix:survey_questionnaire}
\begingroup
\renewcommand{\arraystretch}{1.0} % Adjust row spacing for compactness

% \tiny
\tiny
\begin{longtable}{||p{5cm}|p{5cm}|p{5cm}||} % 3 columns, spanning full width
    \hline
    \textbf{Survey Questions} & \textbf{Options} & \textbf{Answer Frequency} \\
    \hline
    \endfirsthead
    
    \hline
    \textbf{Survey Questions} & \textbf{Options} & \textbf{Answer Frequency} \\
    \hline
    \endhead
        
    % First Question
    \multirow{3}{5cm}{How do you collect the evidence (logs, traces, etc.) that is used as the basis to generate CTI?} 
    & 1. Logs and traces are collected on an active network as part of incident forensics & 1. 9 \\ 
    & 2. Logs and traces are collected in dedicated analysis environments (e.g., detonation of malware in sandboxes) & 2. 17 \\ 
    & 3. Other (text) & 3. i) Online open-source tooling exposed to users, which used the tools and the collection of data submitted is used to generate or validate CTI. \\ 
    & & ii) We collect logs from various SaaS services. \\ 
    & & iii) Logs sent to SIEM solutions. \\ 
    \hline
    
    % Second Question
    \multirow{6}{5cm}{What is your main approach in analyzing logs to generate CTI?} 
    & 1. Formal methods (graph analysis, tree-based parsing) & 1. 4 \\ 
    & 2. Scripted Data Parsing (analyzing a CSV or analyzing a Wireshark dump using Python) & 2. 15 \\ 
    & 3. Machine learning (decision trees, regression, classification...) & 3. 5 \\ 
    & 4. LLMs & 4. 3 \\ 
    & 5. Manual methods & 5. 13 \\ 
    & 6. Other (text) & 6. i) Statistics (outside machine learning). \\    
    \hline
    
    % Third Question
    \multirow{6}{5cm}{What are the issues that you encounter while generating CTI from logs?} 
    & 1. Noisy logs that prevent the extraction of relevant events & 1. 13\\ 
    & 2. Difficulty identifying IoCs in logs and events & 2. 7\\ 
    & 3. Lack of aggregation and integration of logs from different sources & 3. 9\\
    & 4. Resources necessary to generate human-readable threat reports & 4. 12 \\
    & 5. Correlating events to high-level TTPs (e.g., MITRE techniques) & 5. 10 \\
    & 6. Other (text) & 6.  i) Time it takes, ii) Lack of resources to create necessary parsers to make the information in vast amount of types logs useful.\\
    \hline
    
    % Fourth Question
    \multirow{6}{5cm}{What are opportunities to use AI in the process of generating CTI from logs? (Select all the options that apply)} 
    & 1. Filtering raw, noisy logs to extract relevant events & 1. 15\\ 
    & 2. Identifying IoCs in logs and events & 2. 12\\ 
    & 3. Aggregating logs from different sources & 3. 11\\ 
    & 4. Mapping threat events to TTPs (e.g., MITRE techniques) & 4. 13\\ 
    & 5. Other (text) & 5. i) Processing data stream which were unprocessed previously (e.g. extraction text from audio), improving time correlation and location based on LLM models with key time events\\ 
    & & ii) Creation of diagrams and other visualizations to be used in reports, but based on human input.\\

    % Fifth Question
    \hline
    \multirow{4}{5cm}{What are your primary concerns regarding the use of AI to extract CTI from logs and other forensic evidence? (Select all the options that apply)}
    & 1. Preprocessing the inputs (logs) in a way that they can be consumed by AI tools (e.g., feature engineering) & 1. 6\\
    & 2. Losing context of logs when logs are too long (e.g., LLMs have limited context windows) & 2. 11\\
    & 3. Reliability of AI-based solutions for filtering raw logs (e.g., errors in correlating events, erroneous prioritization of events) & 3. 16\\
    & 4. Other (text) & 4. i) Some unknown bias when the model was built and the description of the model training is not properly documented.\\
    & & ii)Leaking of confidential/classified.
data\\

    % Sixth Question
    \hline
    \multirow{5}{5cm}{What different channels do you use to share CTI? (Select all the options that apply)}
    & 1. Private channels (e.g., emails) & 1. 14\\
    & 2. Institutional channels (e.g., CISA) & 2. 4\\
    & 3. Public channels (e.g., OSINT, blogs, discord, Virustotal) & 3. 7\\
    & 4. Threat intelligence sharing platforms (e.g., MISP) & 4. 10\\
    & 5. Other (text) & 5. Real-time streaming platforms such as CocktailParty. \\
    
    % Seventh Question
    \hline
    \multirow{4}{5cm}{How do you codify your CTI? (Select all the options that apply)}
    & 1. Natural language (unstructured) & 1. 13\\
    & 2. Natural language (structured) & 2. 8\\
    & 3. Machine-consumable formats (e.g., STIX/TAXII) &  3. 5\\
    & 4. Other (text) & 4. Visualizations (images, videos).\\
    \hline

    % Eighth Question
    \multirow{5}{5cm}{What are the biggest problems you face while sharing CTI from both technical and policy perspectives? (Select all the options that apply)}
    & 1. Ensuring regulatory compliance or preventing unlawful behaviors when sharing CTI (e.g., conforming to EU-based regulations) & 1. 6\\
    & 2. Ensuring shared CTI does not expose private information (e.g., with respect to a company's policies) & 2. 17\\
    & 3. Translating CTI into different formats for different sharing feeds & 3. 7\\
    & 4. Ownership Compliance (Agency A shares CTI with Agency B, but Agency A does not want Agency B not to share data further) & 4. 6\\
    & 5. Other (text)  & 5. i) Lack of standardization in reporting and formats. Even with STIX. Practitioners don't always use the same contextualizations (taxonomies) or standards, and even if they do there are often different understandings of data.\\
    \hline

    % Tenth Row
    
    \multirow{4}{5cm}{What are some promising opportunities for AI in CTI sharing? (Select all the options that apply)}
    & 1. Ensuring regulatory compliance for CTI before sharing & 1. 7 \\
    & 2. Finding private information and removing/masking it from CTI & 2. 14\\
    & 3. Translating CTI into different formats for various sharing feeds & 3. 11\\
    & 4. Other (text) & 4. i) Automating the analysis large dataset which were unprocessed previously.\\
    && ii) Summarization, labeling of data. \\
    \hline
    What are your primary concerns regarding the use of AI in threat intelligence sharing? (Select all the options that apply)
    & 1. Loss of context in threat data due to the limited capabilities of available AI systems & 1. 6 \\  
    & 2. Insufficient controls that might bring AI-based systems to break privacy or regulatory compliance & 2. 12\\
    & 3. Lack of confidence in the ability of AI-based systems to correctly carry out a task & 3. 14\\
    & 4. AI-based systems might introduce inaccuracies in the data/results & 4. 13\\
    & 5. Other (text) & 5. Leaking of confidential/classified data.\\
    \hline
    
    What are the biggest hurdles of incorporating AI in practitioners' workflow? (Select all the options that apply)
    & 1. Cost of using AI & 1. 6\\
    & 2. Risk of unreliable outputs & 2. 15\\
    & 3. Data privacy concerns & 3. 13\\
    & 4. Need for explainability & 4. 8\\
    & 5. Problems with maintainability & 5. 8\\
    & 6. Lack of expertise & 6. 7\\
    & 7. Other (text) & 7. The lack of real open-source AI models.\\
    \hline
    \caption{ Survey questionnaire and practitioners' responses to understand CTI workflows, challenges, and AI-driven opportunities.}
    \label{tab:survey_questions}
\end{longtable}
\endgroup

\end{document}